\documentclass[aps,prx,fixfloat,twocolumn]{revtex4-2}
\usepackage{graphicx,amsmath,amsfonts,bm, color}
\usepackage{wasysym}

\usepackage{xcolor,hyperref}
\hypersetup{
   colorlinks,
   linkcolor={blue!50!black},
   citecolor={blue!50!black},
   urlcolor={blue!80!black}
}

\usepackage{enumitem}

\renewcommand{\=}{\!=\!}

\DeclareMathAlphabet{\mathitbf}{OML}{cmm}{b}{it}

\begin{document}

\title{Quenched disorder and instability control dynamic fracture in three dimensions}
\author{Yuri Lubomirsky}
\author{Eran Bouchbinder}
\affiliation{Chemical and Biological Physics Department, Weizmann Institute of Science, Rehovot 7610001, Israel}

\begin{abstract}
Materials failure in 3D still poses basic challenges. We study 3D brittle crack dynamics using a phase-field approach, where Gaussian quenched disorder in the fracture energy is incorporated. Disorder is characterized by a correlation length $R$ and strength $\sigma$. We find that the mean crack velocity $v$ is bounded by a limiting velocity, which is smaller than the homogeneous material's prediction and decreases with $\sigma$. It emerges from a dynamic renormalization of the fracture energy with increasing crack driving force $G$, resembling a critical point, due to an interplay between a 2D branching instability and disorder. At small $G$, the probability of localized branching on a scale $R$ is super-exponentially small. With increasing $G$ this probability quickly increases, leading to misty fracture surfaces, yet the associated extra dissipation remains small. As $G$ is further increased, branching-related lengthscales become dynamic and persistently increase, leading to hackle-like structures and to a macroscopic contribution to the fracture surface. The latter dynamically renormalizes the actual fracture energy until eventually any increase in $G$ is balanced by extra fracture surface, with no accompanying increase in $v$. Finally, branching width reaches the system's thickness such that 2D symmetry is statistically restored. Our findings are consistent with a broad range of experimental observations.
\end{abstract}

\maketitle

Materials failure is a physical phenomenon of prime scientific and technological importance, which continues to pose fundamental challenges~\cite{ravi1998dynamic,fineberg.99,bonamy.11,bouchbinder.14,fineberg2015recent}. It is mediated by the propagation of cracks, which are dynamic defects that feature extreme conditions near their edges --- mimicking a mathematical singularity --- and accelerate to relativistic velocities (relative to elastic wave-speeds) in brittle materials~\cite{freund,99bro}. Moreover, crack dynamics involve strongly non-equilibrium physics, long-range elastodynamic interactions and a wide range of spatiotemporal scales, which still resist a complete theoretical understanding.

Things somewhat simplify in 2D, i.e., when experiments in thin samples are considered~\cite{livne.07,lubomirsky2018}, where a crack is the trajectory left behind a propagating tip (a point). Major progress in understanding rapid crack dynamics in 2D has been recently made~\cite{lubomirsky2018,bouchbinder2008,bouchbinder.09b,chen2017,vasudevan2021oscillatory}. In particular, high-velocity symmetry-breaking instabilities have been predicted, in quantitative agreement with experiments, including the 2D oscillatory instability~\cite{livne.07} and the 2D branching instability (sometimes also termed macro-branching/tip-splitting)~\cite{lubomirsky2018}. These two instabilities are linear instabilities, which are spontaneously triggered by infinitesimal noise, once the critical conditions are met.

\begin{figure}[h]
\center
\includegraphics[width=0.42\textwidth]{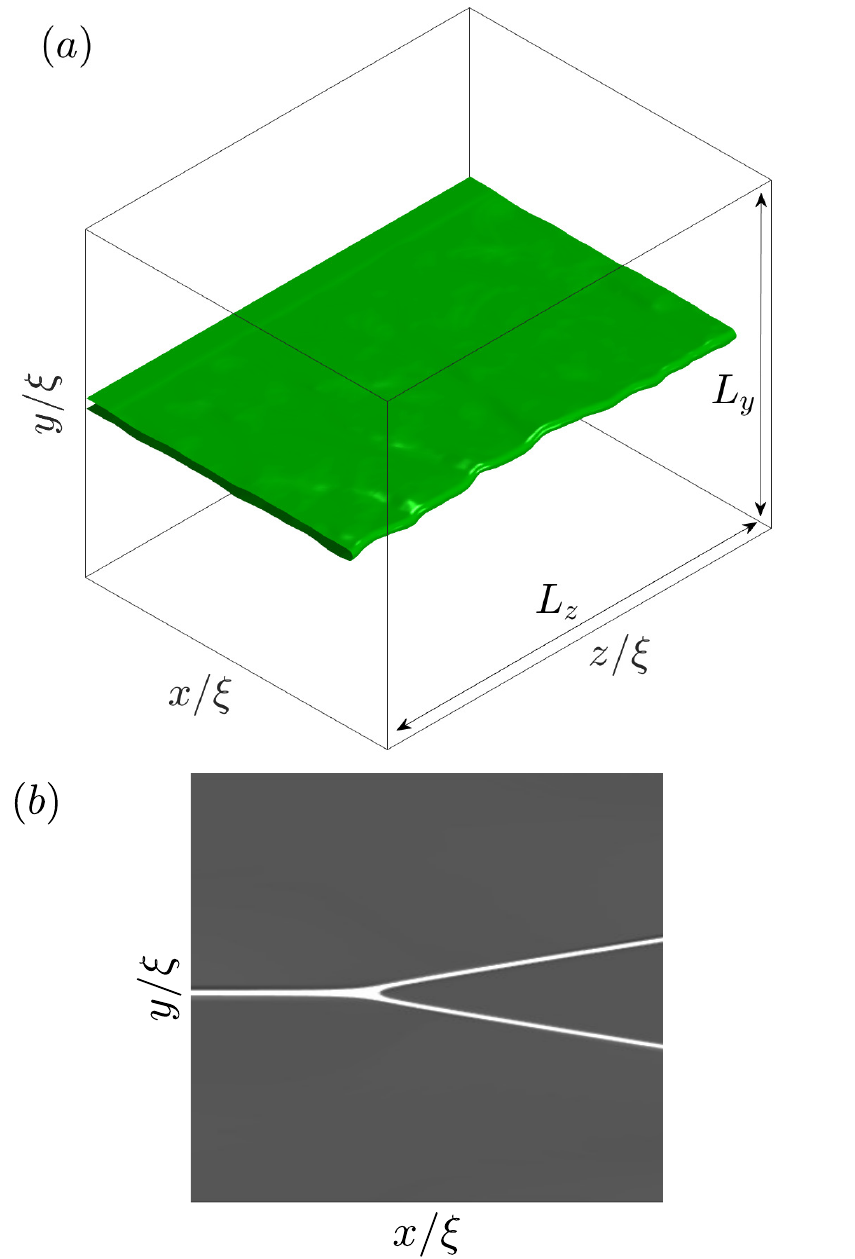}
\caption{(a) A sketch of a 3D tensile crack experiment in a long bar of height $L_y$ and thickness $L_z$. The long bar features a length $L_x\!\gg\!L_y,L_z$, and only a small section of it is shown. The crack is located in the middle (symmetry) plane and will subsequently propagate predominantly  in the $x$ direction in response to a sufficiently large crack driving force $G$ (related to the tensile loading, applied in the $y$ direction). We employ periodic boundary conditions along the $z$ direction~\cite{SI}. The coordinates are normalized by the dissipation length $\xi$, see text for definition. (b) A linear crack branching instability in 2D, i.e., $L_z\!=\!0$, occurring at a very high propagation velocity $v_{_{\rm B}}\!=\!0.95c_{_{\rm R}}$ ($c_{_{\rm R}}$ is the Rayleigh wave-speed), once a threshold tensile driving force $G_{\rm B}$ is surpassed. See text and Fig.~\ref{fig:fig2} for additional discussion.}
\label{fig:fig1}
\end{figure}

Our corresponding understanding of dynamic fracture in 3D lags far behind. In 3D, the sample thickness is no longer a negligible dimension and a crack is the surface left behind a propagating line, i.e., here the crack edge is a 2D front, see Fig.~\ref{fig:fig1}a. While available experiments on brittle amorphous materials shed important light on the phenomenology of dynamic fracture in 3D~\cite{ravi1984experimental_II,ravi1984experimental_III,scheibert2010brittle,guerra2012understanding,tanaka1998discontinuous,steps2017,baumberger.08,sharon1996microbranching,sharon1998universal,sharon1999dynamics,livne2005universality,lawn,johnson1968microstructure,rabinovitch2000origin,jiao2015macroscopic}, our fundamental understanding of the underlying physics remains limited. These experiments revealed a variety of 3D out-of-plane crack structures, including micro-cracking damage~\cite{ravi1984experimental_II,ravi1984experimental_III,scheibert2010brittle,guerra2012understanding}, surface steps~\cite{tanaka1998discontinuous,steps2017}, cross-hatching patterns~\cite{baumberger.08}, micro-branches~\cite{sharon1996microbranching,sharon1998universal,sharon1999dynamics,livne2005universality} and mirror-mist-hackle patterns~\cite{lawn,johnson1968microstructure,rabinovitch2000origin,jiao2015macroscopic}, all emerging at mean crack front velocities smaller than the 2D instability thresholds.

Here, we study 3D dynamic fracture using a flexible computational framework in which material quenched disorder is incorporated. It is based on a phase-field fracture approach~\cite{lubomirsky2018,chen2017,vasudevan2021oscillatory,karma2001phase,Karma2004,Hakim.09,das2023dynamics,bleyer2017microbranching,henry2013fractographic}, which is particularly suitable for studying 3D dynamic fracture as it allows cracks to self-consistently select complex 3D trajectories (including topological changes) without imposing any external path selection criteria. Moreover, it allows to track the in silico real-time 3D spatiotemporal dynamics of cracks in a way that goes well beyond current experiments. This framework quantitatively predicted the 2D high-velocity oscillatory and branching instabilities in homogeneous materials~\cite{lubomirsky2018,chen2017,vasudevan2021oscillatory}, as well as the dynamics of crack front waves in 3D homogeneous materials with isolated heterogeneities~\cite{das2023dynamics}.

Evidence regarding 3D dynamic fracture is largely obtained in experiments on amorphous materials, such as various glassy polymers~\cite{ravi1984experimental_II,ravi1984experimental_III,scheibert2010brittle,guerra2012understanding,sharon1996microbranching,sharon1998universal,sharon1999dynamics,jiao2015macroscopic}, silica glasses~\cite{sharon1998universal,lawn,johnson1968microstructure,rabinovitch2000origin} and brittle elastomers~\cite{livne2005universality}. These materials are intrinsically disordered, featuring fluctuations in material properties over various lengthscales. Material disorder and heterogeneity have been extensively discussed mainly in the context of in-plane fracture roughness in 3D and mostly in the quasi-static regime (e.g.,~\cite{bonamy.11,Ponson2023}). Some recent studies considered the effect of material heterogeneity on dynamic in-plane cracks~\cite{roch2023dynamic} and on quasi-static out-of-plane cracks~\cite{lebihain2020effective}. Yet, the interaction of a 3D elastodynamic crack with continuous quenched disorder has not been studied theoretically. We show that the interplay between quenched disorder and the 2D branching instability controls dynamic fracture in 3D, in agreement with a wide range of experimental observations.\\

\newpage
\hspace{-0.35cm}{\large \bf Results}\\

\hspace{-0.35cm}{\bf{\em Dimensionality, stability and disorder-induced limiting velocity}}\\

A canonical experiment for probing failure dynamics involves an initial planar crack located in the middle plane of a long 3D rectangular bar, see Fig.~\ref{fig:fig1}a. The bar features height $L_y$, thickness $L_z$ and length $L_x\!\gg\!L_y,L_z$, where the $(x,y,z)$ coordinate system is defined in Fig.~\ref{fig:fig1}. It is loaded symmetrically by small tensile displacements $u_y(x,y\!=\!\pm L_y/2,z,t)\=\delta/2$, where $\bm{u}(x,y,z,t)$ is the 3D displacement field and $t$ is time. The tensile loading corresponds to a crack driving force $G\!\sim\!E\delta^2/L_y$ (the linear elastic energy per unit area stored far ahead of the initial crack)~\cite{freund,99bro}, where $E$ is Young's modulus. Once $G$ surpasses the minimal quasi-static fracture energy $\Gamma_0$, the tensile (mode-I) crack front starts propagating predominantly in the $x$ direction. Understanding the subsequent spatiotemporal dynamics as a function of $G\!>\!\Gamma_0$ remains a major open challenge.

Experiments are limited in probing the real-time spatiotemporal dynamics of cracks in 3D, mainly because continuous imaging of extended dynamic defects that propagate at high velocities inside a material is currently not available. Consequently, large-scale computer simulations offer a powerful complementary approach. The phase-field approach, mentioned above, allows to numerically solve the 3D problem delineated in Fig.~\ref{fig:fig1}a, while accurately probing the in silico real-time 3D spatiotemporal dynamics of the crack. This is achieved by solving the linear elastodynamic field equations for $\bm{u}(x,y,z,t)$ inside the bar, coupled to an auxiliary field --- the scalar phase-field $\phi(x,y,z,t)$~\cite{lubomirsky2018,chen2017,vasudevan2021oscillatory,das2023dynamics,SI}. The latter satisfies its own dissipative field equation, which features a characteristic dissipation length $\xi$ and a dissipation time $\tau$~\cite{lubomirsky2018,chen2017,vasudevan2021oscillatory,das2023dynamics,SI}.

In the presence of the intense, nearly singular deformation fields in the vicinity of the moving crack front, $\phi(x,y,z,t)$ both spontaneously generates the traction-free surfaces that define the crack~\cite{lubomirsky2018,chen2017,vasudevan2021oscillatory,das2023dynamics,SI} --- and hence self-consistently selects the crack trajectory --- and gives rise to a rate-dependent fracture energy $\Gamma(v)$ controlled by the dissipation time $\tau$, where $v$ is the crack velocity (note that $\Gamma(v\!\to\!0)\=\Gamma_0$). Another major advantage of this computational approach, essential for the present work, is the ability to tune material properties in a controlled manner in ways that are difficult to obtain in the laboratory. The phase-field fracture approach played a central role in recent progress in understanding rapid crack dynamics in 2D~\cite{lubomirsky2018,chen2017,vasudevan2021oscillatory}. The latter corresponds to taking the $L_z/\xi\!\to\!0$ limit in the 3D setup of Fig.~\ref{fig:fig1}a, leading to a long strip configuration~\cite{freund,99bro}.

\begin{figure}[ht!]
\center
\includegraphics[width=0.495\textwidth]{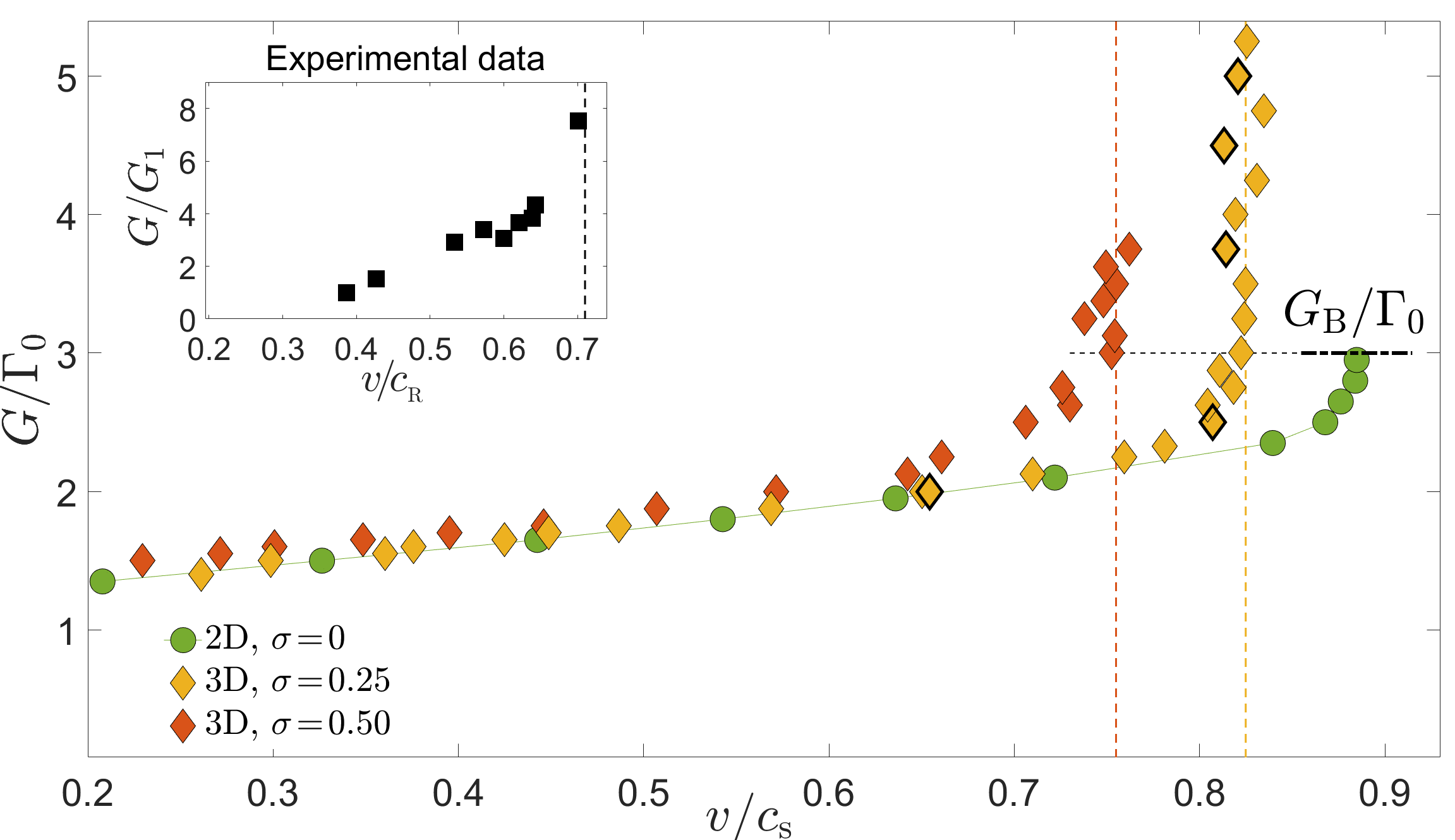}
\caption{The crack driving force $G/\Gamma_0$ vs.~steady-state crack velocity $v/c_{\rm s}$ for a 2D homogeneous material (green circle) and two 3D heterogeneous materials with $\sigma\!>\!0$ (yellow diamonds for $\sigma\!=\!0.25$ and brown diamonds for $\sigma\!=\!0.50$), obtained in phase-field simulations~\cite{SI}. The thick-boundary diamonds correspond to the 5 snapshots shown in Fig.~\ref{fig:fig3}. Steady-state solutions in 2D homogeneous materials do not exist for $G\!>\!G_{\rm B}$ (horizontal dashed-dotted line), upon which a linear branching instability sets in (see Fig.~\ref{fig:fig1}b). In 3D materials, the average steady-state crack velocity is bounded by a $\sigma$-dependent limiting velocity $v_{_{\rm lim}}(\sigma)$ (vertical dashed lines), where $G$ increases in a critical-like manner.
(inset) $G$ (normalized by the first data point $G_1$) vs.~steady-state crack velocity $v/c_{\rm R}$ measured in experiments on a glassy polymer (Polymethyl Methacrylate, PMMA), data extracted from Fig.~5 of~\cite{sharon1996microbranching}. The vertical dashed line, highlighting the similarity to the simulational data in the main panel, is added as a guide to the eye.}
\label{fig:fig2}
\end{figure}

In Fig.~\ref{fig:fig2}, we plot (green circles) the imposed crack driving force $G/\Gamma_0$ against the resulting steady-state crack velocity $v$ (normalized by the shear wave-speed $c_{\rm s}$) obtained in 2D phase-field calculations in a homogeneous material~\cite{vasudevan2021oscillatory}. It is observed that $G$ is a monotonically increasing function of $v$, where each point along the curve corresponds to a straight crack that respects the global tensile (mode-I) symmetry of the system. Moreover, the curve is linear up to a velocity close to the Rayleigh wave-speed, here $c_{_{\rm R}}\!=\!0.93c_{\rm s}$, which is predicted to be the upper bound on the velocity of cracks based on Linear Elastic Fracture Mechanics (LEFM) theory, in the absence of symmetry-breaking instabilities~\cite{freund,99bro}. Beyond this regime, $G$ strongly increases with $v$ due to crack tip blunting, leading to increased dissipation~\cite{lubomirsky2018,vasudevan2021oscillatory}.

The 2D $G$ vs.~$v$ curve terminates at $G_{\rm B}$ (horizontal dashed-dotted line), at a velocity very close to ---  yet slightly smaller than --- $c_{_{\rm R}}$, upon which a branching instability (sometimes termed macro-branching/tip-splitting) takes place. That is, a straight tensile crack branches/splits into two coexisting cracks~\cite{lubomirsky2018,vasudevan2021oscillatory}, shown in Fig.~\ref{fig:fig1}b, providing a mechanism for excess dissipation due to additional fracture surfaces~\cite{Eshelby1970,adda-bedia.2007}. For $G\!>\!G_{\rm B}$, steady-state solutions no longer exist in 2D. The high-velocity 2D branching instability, previously demonstrated in~\cite{lubomirsky2018,vasudevan2021oscillatory}, has been experimentally observed in quasi-2D brittle materials~\cite{ravi1984experimental_III,lubomirsky2018,jiao2015macroscopic,sharon1999dynamics,kobayashi1974crack,sun2021state,yoon2022situ}. Finally, since in steady-state $G$ is balanced by the fracture energy, we identify the $G$ vs.~$v$ curve as $\Gamma_{\!_{\rm 2D}}(v)$ and hence the 2D branching velocity $v_{_{\rm B}}$ corresponds to $\Gamma_{\!_{\rm 2D}}(v_{_{\rm B}})\=G_{\rm B}$. We note in passing that the 2D oscillatory instability, which emerges at a velocity slightly smaller than $v_{_{\rm B}}$~\cite{chen2017}, does not emerge here~\cite{SI}.

Branching in 2D exists in purely homogeneous materials and hence is a linear instability. We next considered the effect of dimensionality on the stability of cracks in homogeneous materials by studying 3D systems with a finite thickness $L_z$, as in Fig.~\ref{fig:fig1}a, driven by $G\!<\!G_{\rm B}$. We found that tensile cracks in our 3D homogeneous system are remarkably stable, both linearly and nonlinearly. In particular, for a wide variety of crack shape perturbations (of different spatial extent and amplitudes) as well as for a wide range of isolated heterogeneities (of different sizes, shapes and locations), cracks transiently go out of their initial plane, but then decay back to the tensile symmetry plane. That is, no persistent 3D out-of-plane crack instabilities exist in homogeneous materials for $G\!<\!G_{\rm B}$, hence for $v\!<\!v_{_{\rm B}}$ (persistent in-plane crack front waves are triggered, as discussed very recently in~\cite{das2023dynamics}). This is in sharp and qualitative disagreement with a wide range of experimental observations that show that tensile cracks in 3D brittle amorphous materials reveal persistent out-of-plane structures at velocities smaller than $v_{_{\rm B}}$, as already noted above.

What is the origin of this understanding gap? What is the crucial physical ingredient missing? In order to gain insight into these fundamental questions, we stress that the experimental evidence at hand was obtained for amorphous materials, which are homogeneous at macroscopic scales, but feature disorder at smaller scales. The quantitative characterization and understanding of disordered materials is still incomplete; in particular, while it is known that various material properties (e.g., elastic moduli~\cite{wagner2011local,kapteijns2021elastic}) feature significant fluctuations over supermolecular lengthscales, we do not yet have robust and quantitative knowledge of the statistics of fundamental physical quantities such as the fracture energy. Consequently, in order to test the possible importance of material quenched disorder, our strategy would be to adopt minimal assumptions about material quenched disorder and explore its implications for bridging the qualitative gap in our understanding of 3D dynamic fracture.

Specifically, we introduce quenched disorder in the form of a Gaussian distribution in the dimensionless quasi-static ($v\!\to\!0$) fracture energy $\bar{\Gamma}/\Gamma_0$, with unity mean, standard deviation determined by $\sigma$ and spatial correlation length $R$~\cite{SI}. We hereafter set $R\=10\xi$ and study the effect of $\sigma$ on 3D dynamic fracture. In Fig.~\ref{fig:fig2}, we plot (diamonds) $G/\Gamma_0$ against the average steady-state crack velocity $v/c_{\rm s}$ obtained in 3D phase-field calculations as in Fig.~\ref{fig:fig1}a, for two values of $\sigma$. Recall that in steady state, energy balance implies $\Gamma(v)\=G$~\cite{freund,99bro}, where $\Gamma(v)$ is the actual fracture energy. It is observed that $\Gamma(v)\=G$ closely follows the 2D homogeneous material curve at relatively small $v$, then it deviates from it at larger $v$'s until it reaches a limiting (terminal) velocity $v_{_{\rm lim}}(\sigma)\!<\!v_{_{\rm B}}\!<\!c_{_{\rm R}}$, consistently with numerous experiments (e.g.,~\cite{fineberg.99,sharon1996microbranching} and the compilation of experimental works in Table.~1 of~\cite{ravi1984experimental_III}). Moreover, $d\Gamma/dv$ appears to diverge for $v\!\to\!v_{_{\rm lim}}$, also consistently with experiments~\cite{sharon1996microbranching}, cf.~inset of Fig.~\ref{fig:fig2}.

The results in Fig.~\ref{fig:fig2} can be described by the steady-state energy balance
\begin{equation}
\Gamma(v)=\Gamma_{\!_{\rm 2D}}(v)+\delta\Gamma_{\!_{\rm 3D}}(v,\sigma)=G \ ,
\label{eq:energy_balance}
\end{equation}
where $\delta\Gamma_{\!_{\rm 3D}}(v,\sigma)$ corresponds to 3D disorder-induced corrections to the 2D homogeneous-material $\Gamma_{\!_{\rm 2D}}(v)$. We note that we do not explicitly incorporate the dependence of $\delta\Gamma_{\!_{\rm 3D}}(v,\sigma)$ on the correlation length $R$ since it is kept constant, as stated above. For relatively small velocities, we have $\delta\Gamma_{\!_{\rm 3D}}(v,\sigma)\!\ll\!\Gamma_{\!_{\rm 2D}}(v)\!\simeq\!\Gamma(v)$ and the behavior essentially identifies with that of 2D homogeneous materials. At some $\sigma$-dependent characteristic $v$, $\delta\Gamma_{\!_{\rm 3D}}(v,\sigma)$ is no longer negligible compared to $\Gamma_{\!_{\rm 2D}}(v)$, i.e., 3D disorder-induced effects become important. Finally, as $v\!\to\!v_{_{\rm lim}}$, $\delta\Gamma_{\!_{\rm 3D}}(v,\sigma)$ becomes sizable/dominant and appears to feature a critical behavior, i.e., $d\Gamma(v)/dv$ appears to diverge. Our main challenge in the remaining of the paper is to understand the physics behind $\delta\Gamma_{\!_{\rm 3D}}(v,\sigma)$.\\
\begin{figure*}[ht!]
\center
\includegraphics[width=1\textwidth]{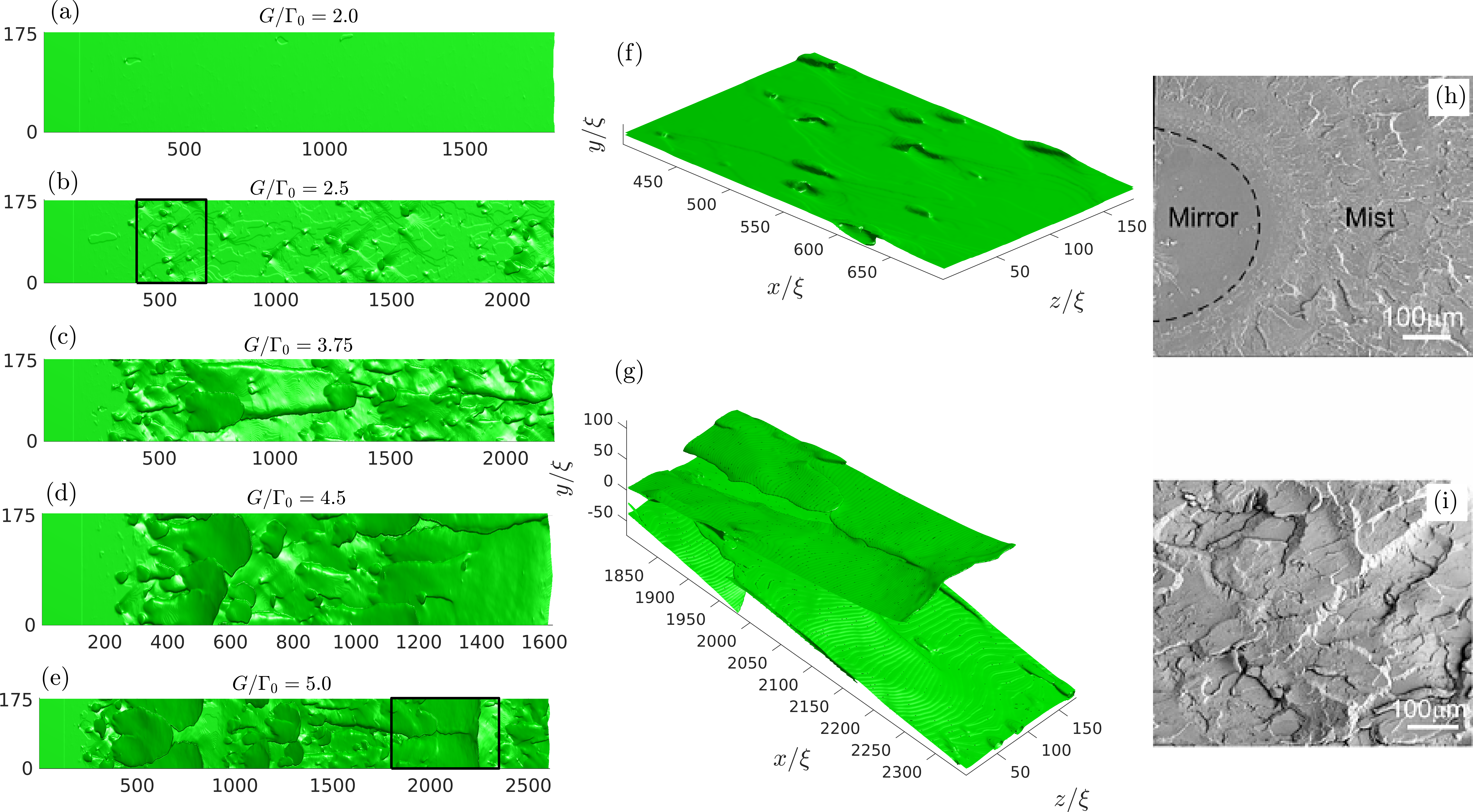}
\caption{(left column, panels a-e) Uppermost out-of-plane crack structures --- corresponding to the phase-field $\phi({\bm x})\!=\!1/2$ iso-surface~\cite{SI} --- inside the bulk (top view), for the five $G$ values indicated (corresponding to the thick-boundary diamonds in Fig.~\ref{fig:fig2}), see text for extensive discussion. In each panel, the long axis is the propagation ($x$) direction and the perpendicular one in the crack front direction ($z$ axis). Both axes are expressed in units of $\xi$, and the basic length along $z$ is twice that of along $x$, for visual clarity. The initial crack front is marked by the bright line in the left part of each panel. (middle column, panels f-g) Tilted zoom-in 3D view of the regions marked by the rectangles in panels (b) and (e), respectively. See text for discussion. (right column, panels i-h) Fracture surface morphologies (fractography) measured in dynamic fracture experiments in a glassy polymer (PMMA) involving accelerating 3D cracks (propagating from left to right), obtained through SEM imaging~\cite{jiao2015macroscopic}. Panel (h) shows the mirror-to-mist transition, where the mist region is characterized by rather isolated, small-amplitude out-of-plane surface structures (adapted from Fig.~4b of~\cite{jiao2015macroscopic}). Initially, the crack does not capture the entire sample thickness (vertical axis in the image) and it propagates predominantly radially. Panel (h) bears qualitative similarities to panels (b) and (f). Panel (i) shows the hackle region, appearing at larger crack propagation distances in the same experiment (adapted from Fig.~4d of~\cite{jiao2015macroscopic}). The observed wider and elongated out-of-place surface structures bear clear similarities to panels (c) and (d).}
\label{fig:fig3}
\end{figure*}

\hspace{-0.35cm}{\bf{\em Out-of-plane crack structures and the localized branching instability}}\\

To start addressing this challenge, we focus on out-of-plane crack structures that correspond to the 3D results in Fig.~\ref{fig:fig2} with $\sigma\=0.25$ (to be considered hereafter). Since the crack front experiences both in- and out-of-plane fluctuations, the quantity $v$ hereafter stands for the averaged front velocity~\cite{SI}. We take advantage of our computational platform and consider out-of-plane crack structures inside the bulk of the material, not just their fractographic signature (i.e., the patterns left on the postmortem fracture surface). In Fig.~\ref{fig:fig3} (left column), we present a sequence of out-of-plane crack structures as a function of the crack driving force $G$, covering the various regimes of interest (the data points used are marked by thick boundaries in Fig.~\ref{fig:fig2}).

Figure~\ref{fig:fig3}a corresponds to $G/\Gamma_0\=2$, for which $\delta\Gamma_{\!_{\rm 3D}}(v,\sigma)\!\ll\!\Gamma_{\!_{\rm 2D}}(v)$. It is observed that out-of-plane protrusions are very rare and small, i.e., the crack is essentially planar (mirror-like), which is consistent with the agreement between $\Gamma(v)$ and $\Gamma_{\!_{\rm 2D}}(v)$ in this regime. Figure~\ref{fig:fig3}b corresponds to $G/\Gamma_0\=2.5$, where the deviation of $\Gamma(v)$ from $\Gamma_{\!_{\rm 2D}}(v)$ is still small. It is observed that out-of-plane structures, of width $\Delta{z}$ and length $\Delta{x}$ (marked in Fig.~S3 in~\cite{SI}), emerge at a significantly higher probability compared to panel (a). The out-of-plane structures, leading to a misty fracture surface, are mainly localized branching events of characteristic width comparable to the disorder correlation length $R\=10\xi$, as highlighted in the tilted zoom-in view in Fig.~\ref{fig:fig3}f. The out-of-plane profile of asymmetric localized branches (i.e., when the up-down symmetry is broken) is consistent with experiments, see Fig.~S4 in~\cite{SI}. In addition, each localized branching event is accompanied by V-shaped tracks (clearly observed in Fig.~\ref{fig:fig3}f), which are reminiscent of typical tracks left by crack front waves~\cite{sharon2002crack,fineberg2003crack,adda-bedia.13,das2023dynamics}. Indeed, the propagation velocity of these tracks is consistent with that of crack front waves, see Fig.~S3 in~\cite{SI}.

The character of the out-of-plane structures appearing in Fig.~\ref{fig:fig3}c, corresponding to $G/\Gamma_0\=3.75$, qualitatively changes. The localized branching events therein become significantly wider and longer, i.e., leading to hackle-like structures featuring larger $\Delta{z}$ and $\Delta{x}$, hence larger areas. With increasing $G$, as in Figs.~\ref{fig:fig3}d-e, $\Delta{z}$ and $\Delta{x}$ further increase. In Fig.~\ref{fig:fig3}e, the width $\Delta{z}$ becomes (statistically) comparable to the system thickness $L_z$, which implies that 2D symmetry is statistically restored. This 3D-to-2D transition is manifested in branching events that are no longer localized, i.e., can be regarded as macro-branches~\cite{ravi1984experimental_III,sharon1999dynamics,kobayashi1974crack,bleyer2017microbranching,henry2013fractographic} since they capture the entire system thickness and propagate sizable distances, as highlighted in the tilted zoom-in view in Fig.~\ref{fig:fig3}g. Figures~\ref{fig:fig3}c-e correspond to $G$ values for which the crack reached the limiting velocity, $v\!\to\!v_{_{\rm lim}}$, where $\delta\Gamma_{\!_{\rm 3D}}(v,\sigma)$ makes a sizable/dominant contribution to $\Gamma(v)$ and $d\Gamma(v)/dv$ appears to diverge.
\begin{figure*}[ht!]
\center
\includegraphics[width=1\textwidth]{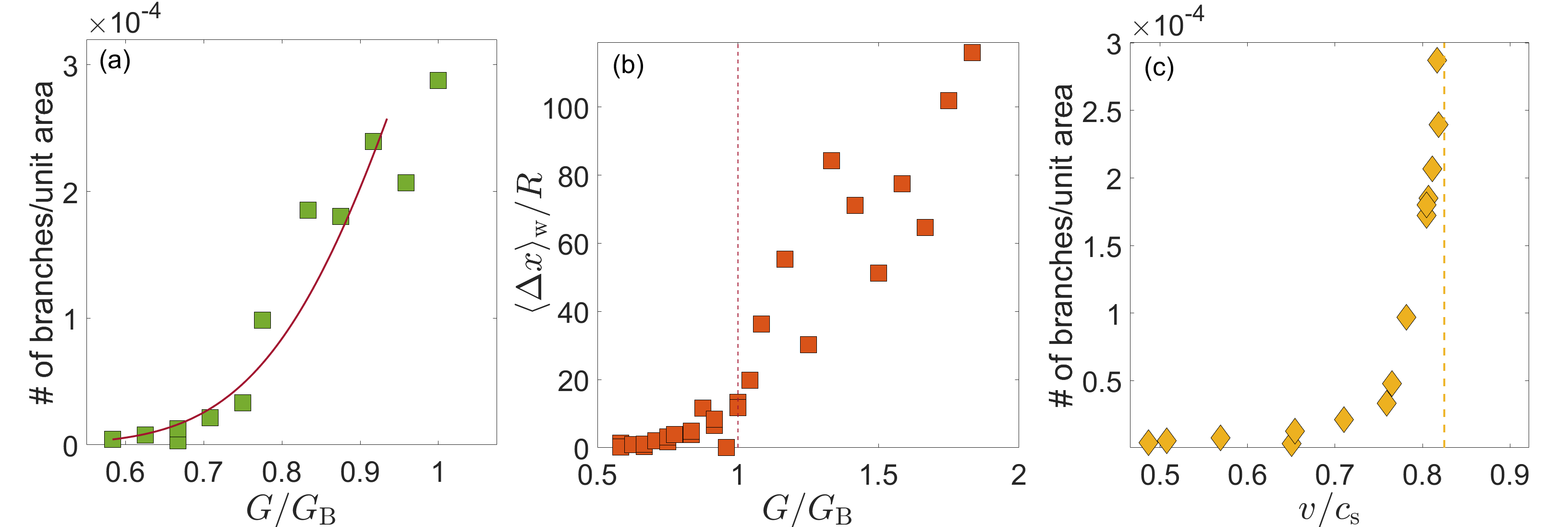}
\caption{ (a) The number of branches per unit planar area as a function of $G/G_{\rm B}\!\le\!1$ (squares). The solid line corresponds to the prediction in Eq.~\eqref{eq:branching_probability}, where $\sigma_{\!_{\rm R}}\!=\!0.233$ is independently predicted~\cite{SI} and the overall amplitude is set to agree with the first data point. (b) The normalized weighted branch length $\langle\Delta{x}\rangle_{\rm w}/R$ (see text for definition) vs.~$G/G_{\rm B}$. The vertical dashed line marks $G\!=\!G_{\rm B}$. (c) The same as panel (a), but vs.~$v$. The vertical dashed line corresponds to $v_{_{\rm lim}}$, as in Fig.~\ref{fig:fig2}.}
\label{fig:fig4}
\end{figure*}

The probabilistic crossover between Fig.~\ref{fig:fig3}a and Fig.~\ref{fig:fig3}b closely resembles the widely-observed mirror-to-mist transition in brittle amorphous materials~\cite{lawn,johnson1968microstructure,rabinovitch2000origin,jiao2015macroscopic}, see Fig.~\ref{fig:fig3}h. The transition in out-of-plane crack structures observed between Fig.~\ref{fig:fig3}b and Fig.~\ref{fig:fig3}c bears strong similarity to the widely documented mist-to-hackle transition, see Fig.~\ref{fig:fig3}i, observed in many brittle amorphous materials, including glassy polymers (e.g.,~\cite{jiao2015macroscopic}) and silica glasses (e.g.,~\cite{rabinovitch2000origin}). Moreover, the 3D-to-2D transition, which is accompanied by macroscopic crack branching and that follows the hackle regime with increasing crack driving force $G$, is also well documented in the very same materials (e.g.,~\cite{ravi1984experimental_III,sharon1999dynamics,rabinovitch2000origin,jiao2015macroscopic}). The correspondence between our findings and a wide range of experimental observations will be further discussed once we gain deeper understanding of the former.\\

\hspace{-0.35cm}{\bf{\em A physical picture and supporting numerics}}\\

Our first goal is to understand the emergence of the localized branching instability in 3D, along with its statistical and dynamical properties. The 2D branching instability in homogeneous materials, cf.~Figs.~\ref{fig:fig1}a and~\ref{fig:fig2}, is triggered at a critical driving force $G_{\rm B}/\Gamma_0$ (see horizontal dashed-dotted line in Fig.~\ref{fig:fig2}). What is the relevance of this instability in 3D in the presence of disorder?

On the scale of the correlation scale $R$, the disordered material can be viewed as locally 2D. Assuming that $G$ in this regime of the 3D dynamics is rather homogeneously distributed along the thickness ($z$ axis), we expect every fluctuation $\bar\Gamma$ in the quasi-static fracture energy that satisfies $G/\bar\Gamma\!>\!G_{\rm B}/\Gamma_0$ to give rise to a localized branching event on a scale $R$. Rearranging the inequality to read $\bar\Gamma/\Gamma_0\!<\!G/G_{\rm B}$ and assuming localized branching events to be independent of one another in this regime such that the Gaussian (normal) distribution of $\bar\Gamma/\Gamma_0$ (with unity mean and standard deviation $\sigma$) can be invoked, we obtain~\cite{SI}
\begin{equation}
p(G;\sigma)\sim 1+\text{erf}\left(\frac{G/G_{\rm B}-1}{\sigma_{\!_{\rm R}}}\right) \ .
\label{eq:branching_probability}
\end{equation}
Here $p(G;\sigma)$ is the localized branching probability, $\text{erf}(\cdot)$ is the error function, which varies super-exponentially with its argument, and $\sigma_{\!_{\rm R}}\!\sim\!\sigma$ is the renormalized strength of disorder on a scale $R$~\cite{SI}.

Equation~\eqref{eq:branching_probability} predicts an extremely small localized branching probability at small $G$ values and a strong increase of the probability with increasing $G$ below $G_{\rm B}$, in qualitative agreement with the results Fig.~\ref{fig:fig3}a-b. To more directly test the prediction in Eq.~\eqref{eq:branching_probability}, we plot in Fig.~\ref{fig:fig4}a the number of localized branches per unit planar area, which is proportional to $p(G;\sigma)$, as a function of $G\!<\!G_{\rm B}$. It is observed that the simple model agrees with the data reasonably well when $G$ is not too close to $G_{\rm B}$.

While 2D branching is a linear instability triggered in purely homogeneous materials (that can be realized in silico) by infinitesimal noise, the corresponding 3D localized branching events with $\bar\Gamma/\Gamma_0\!<\!G/G_{\rm B}$ are finite-disorder instabilities, and hence are non-perturbative with respect to a homogenized material. Moreover, 3D localized branching events with $\bar\Gamma/\Gamma_0\!<\!G/G_{\rm B}\!<\!1$ are expected to be rather compact, roughly of linear scale $R$, and hence short-lived and contributing small excess fracture surface. The reason for this is that while localized branching can be triggered with $G/G_{\rm B}\!<\!1$, it cannot be sustained over scales significantly larger than $R$ (since large local fracture energy fluctuations are likely to be encountered). On the other hand, we expect a change in the dynamics of 3D localized branching events for $G\!>\!G_{\rm B}$, since under these conditions the branching solution has a sizable probability to be sustained over larger distances --- i.e., to feature significantly larger length $\Delta{x}$ --- while co-existing with the planar crack solution.

\begin{figure}[ht!]
\includegraphics[width=0.48\textwidth]{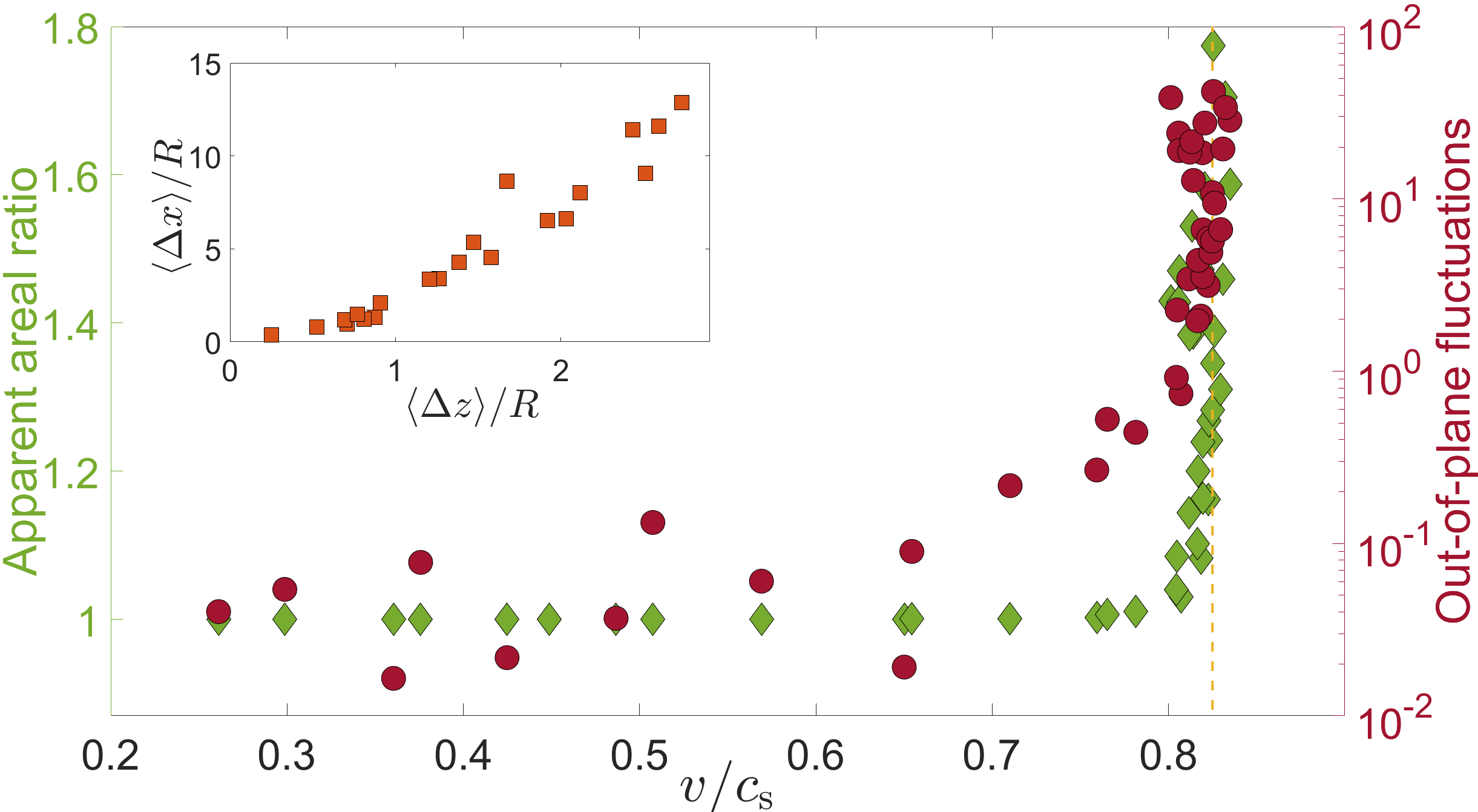}
\caption{The apparent area ratio (see text for definition) vs.~$v/c_{\rm s}$ (green diamonds, left $y$-axis). Out-of-plane fluctuations (see text for definition) in units of $\xi$ vs.~$v/c_{\rm s}$ (brown circles, right $y$-axis, and note the logarithmic scale). The vertical dashed line marks $v_{_{\rm lim}}$. (inset) The averaged branch length $\langle\Delta{x}\rangle/R$ vs.~width $\langle\Delta{z}\rangle/R$.}
\label{fig:fig5}
\end{figure}

To test this expectation, we plot in Fig.~\ref{fig:fig4}b $\langle\Delta{x}\rangle_{\rm w}/R$, the average of $\Delta{x}$ weighted by the branch area (normalized by $R$), against $G/G_{\rm B}$. It is observed that $\langle\Delta{x}\rangle_{\rm w}$ indeed significantly grows with $G$ above $G/G_{\rm B}\!\simeq\!1$ (marked by the vertical dashed line), reaching values much larger than $R$, as expected and in qualitative agreement with Figs.~\ref{fig:fig3}c-e. We note that we consider the averaged branch length weighted by its area because while at smaller $G$ values branching events are independent of one another (cf.~Fig.~\ref{fig:fig3}b), for $G\!>\!G_{\rm B}$ branches effectively ``screen'' the nucleation of other branches and in fact the number of primary branches decreases with increasing $G$ (though secondary structures emerge as well). Moreover, for $G\!>\!G_{\rm B}$ the branch width $\Delta{z}$ is no longer expected to inherit its scale from $R$. Indeed, the results presented in the inset of Fig.~\ref{fig:fig5} indicate a quasi-linear relation between the averaged branch width $\langle\Delta{z}\rangle$ and length $\langle\Delta{x}\rangle$. Fluctuations, to be discussed below, also grow significantly with $G$, eventually corresponding to the statistical 3D-to-2D transition observed in Fig.~\ref{fig:fig3}g. Overall, the results indicate that for $G\!>\!G_{\rm B}$ branching lengthscales become dynamic and a significant extra fracture surface emerges.\\

\hspace{-0.35cm}{\bf{\em The emergence of a limiting velocity and dynamic renormalization of the fracture energy}}\\

We are now in a position to rationalize the steady-state dynamics of 3D cracks as encapsulated in Eq.~\eqref{eq:energy_balance} and in particular the physics underlying $\delta\Gamma_{\!_{\rm 3D}}(v,\sigma)$. The physical picture discussed above and its numerical support indicate that 3D out-of-plane fracture dynamics are related to a branching instability controlled by the driving force $G$ and to quenched disorder, where both the branching probability $p(G;\sigma)$ and the excess fracture surface associated with branching play a role. In particular, at small $G$ the evolution of $p(G;\sigma)$ is dominant, while for $G\!>\!G_{\rm B}$ the extra surface of the branches dominates, with a transition in between.

To relate this picture to the emergence of a limiting velocity $v_{_{\rm lim}}$ and to the properties of $\delta\Gamma_{\!_{\rm 3D}}(v,\sigma)$, we replot in Fig.~\ref{fig:fig4}c the number of localized branches per unit planar area of Fig.~\ref{fig:fig4}a as a function of $v$, i.e., a quantity proportional to $p(v;\sigma)$ instead of $p(G;\sigma)$. It is observed that $p(v;\sigma)$ appears to feature a diverging derivative as $v_{_{\rm lim}}$ is approached (vertical dashed line). That is, the branching probability $p(v;\sigma)$ alone reveals a clear signature of the emergence of a limiting velocity as $G$ approaches $G_{\rm B}$. For $G\!>\!G_{\rm B}$, the branching probability does not further grow (as noted above, it actually decreases), but rather the area of the branches increases such that any increase in $G$ is expected to be balanced by excess fracture surface, without increasing $v$, leading to an increase in $\delta\Gamma_{\!_{\rm 3D}}(v,\sigma)$ at a fixed $v\=v_{_{\rm lim}}$.

To test this expectation, we define the apparent areal ratio as the total area of the uppermost out-of-plane crack structures inside the bulk (as seen in the top view in the left column of Fig.~\ref{fig:fig3}) divided by the nominal (planar) area. The true areal ratio is likely to be larger as secondary structures can develop beneath the uppermost out-of-plane crack structures. The apparent areal ratio is plotted (diamonds, left $y$-axis) in Fig.~\ref{fig:fig5} vs.~$v$ and is observed to be indistinguishable from unity until $v_{_{\rm lim}}$ is approached. At $v\=v_{_{\rm lim}}$, the apparent areal ratio increases in a critical-like manner, as expected. It thus supports the idea that the 3D disorder-induced excess dissipation $\delta\Gamma_{\!_{\rm 3D}}(v,\sigma)$ balances any increase in $G$ at $v\=v_{_{\rm lim}}$ by creating extra fracture surface. That is, the fracture energy $\Gamma(v)$ is dynamically renormalized by the out-of-plane spatiotemporal dynamics of the 3D crack.

Finally, we expect fluctuations in out-of-plane crack structures to significantly increase as $v\!\to\!v_{_{\rm lim}}$. To test this expectation, we extracted for each $(x,y\=0,z)$ point along the original symmetry plane the $y$ difference (in units of $\xi$) between the uppermost out-of-plane crack location (shown on the left column of Fig.~\ref{fig:fig3}) and its lowermost counterpart. We then averaged over all $(x,y\=0,z)$ points and plotted the results in Fig.~\ref{fig:fig5} (circles, right $y$-axis) vs.~$v$ on a log-linear scale. It is observed that the out-of-plane fluctuations are very small for small $v$ (note the logarithmic right $y$-axis), consistently with the branching probability $p(v;\sigma)$ being small in this regime. Around $v/c_{\rm s}\=0.7$, the fluctuations start to grow, yet they still make a negligible contribution to the extra fracture area (see diamonds and left $y$-axis). As $v\!\to\!v_{_{\rm lim}}$, out-of-plane fluctuations grow significantly, featuring the same critical-like behavior as the apparent areal ratio, as expected from the emerging physical picture.\\

\hspace{-0.35cm}{\large \bf Discussion}\\

Our results indicate that 3D dynamic fracture in brittle amorphous materials is controlled by two physical ingredients. The first is a 2D linear instability that is characterized by a crack driving force threshold $G_{\rm B}$, providing a mechanism for generating extra fracture surface, i.e., for increasing the effective fracture energy. The 2D homogeneous material instability involves a topological change, going beyond a single crack, in the form of crack branching. The second ingredient is finite-strength quenched disorder that features a spatial correlation length $R$.

In the presence of finite disorder, the 2D branching instability is transiently excited in 3D at crack driving forces $G$ well below $G_{\rm B}$. It features localization over a scale $R$ that breaks translational invariance along the crack front. Upon increasing $G$ in 3D, the localized branching instability becomes longer-lived and features larger lengthscales for $G\!\simeq\!G_{\rm B}$. The extra fracture surface associated with these increasingly larger branches dynamically renormalizes the effective fracture energy, giving rise to a disorder-dependent limiting velocity $v_{_{\rm lim}}$ that the crack cannot surpass. The emergence of $v_{_{\rm lim}}$ is accompanied by a critical-like behavior in which the fracture energy features a divergent variation with the crack velocity. In this regime, for sufficiently large $G$, branches become so wide that the system statistically recovers translational invariance along the crack front, and branching becomes macroscopic.

This physical picture and its manifestations are consistent with a broad range of experimental observations that were not fully understood previously. At the macroscopic scale, it is widely observed (e.g., in~\cite{fineberg.99,ravi1984experimental_III,sharon1996microbranching}) that cracks in brittle materials attain limiting velocities well below $c_{_{\rm R}}$, the ideal theoretical limit~\cite{freund,99bro}. Moreover, the experiments on steady-state cracks in a glassy polymer --- reproduced in the inset of Fig.~\ref{fig:fig2} --- are consistent with a critical-like behavior of the fracture energy upon approaching $v_{_{\rm lim}}$. At smaller scales, localized instabilities in 3D, mainly in the form of micro-cracking damage~\cite{ravi1984experimental_II,ravi1984experimental_III,scheibert2010brittle,guerra2012understanding} and micro-branching~\cite{sharon1996microbranching,sharon1998universal,sharon1999dynamics,livne2005universality} that involve a topological change and the generation of extra fracture surface, are widely observed.

The localized branching instability extensively discussed above bears close similarity to the micro-branching instability~\cite{sharon1996microbranching,sharon1998universal,sharon1999dynamics,livne2005universality}. The latter features characteristic lengthscales at initiation, most notably minimal micro-branch width and length~\cite{sharon1996microbranching,sharon1998universal,sharon1999dynamics,livne2005universality}, consistently with our findings in which the latter are related to the disorder correlation length $R$. Our results show that localized branching is a probabilistic phenomenon, yet it appears to be rather sharp due to the super-exponential increase of the branching probability at small $G$, which depends on the finite-strength disorder. The probabilistic nature of the micro-branching instability and its dependence on finite disorder are indeed experimentally demonstrated~\cite{livne2005universality,boue2015origin}, in agreement with our findings.

With increasing driving force, micro-branches have been shown to significantly increase in size, to lead to increased effective fracture energy associated with extra fracture surface, to be accompanied by increased out-of-plane fluctuations, to feature a 3D-to-2D transition~\cite{sharon1996microbranching} and eventually to transform into macro-branches~\cite{sharon1999dynamics}, all in agreement with our results. The approximate linear scaling $\langle\Delta{x}\rangle\!\sim\!\langle\Delta{z}\rangle$, cf.~inset of Fig.~\ref{fig:fig5}, is also in agreement with experimental micro-branching observations~\cite{sharon2002crack}. Finally, the out-of-plane profiles of asymmetric localized branches agree with those of experimentally observed micro-branches, as demonstrated in Fig.~S4 in~\cite{SI}.

At intermediate scales, commonly probed by fractography in experiments, our findings are consistent with, and offer novel understanding of, the widely-observed sequence of mirror-mist-hackle-macro-branching morphological transitions~\cite{lawn,johnson1968microstructure,rabinovitch2000origin,jiao2015macroscopic}. This sequence of transitions is commonly observed as a function of crack propagation distance in experiments that involve accelerating cracks. We take advantage of our computational framework, which allows to simulate very long systems~\cite{SI}, to study this sequence of transitions in a controlled manner at the steady-state level, as a function of $G$. Not only our observations, cf.~Fig.~\ref{fig:fig3}, unprecedentedly reproduce this sequence of transitions as $G$ is varied, but we also provided a theoretical understanding of the physics underlying each transition.

The mirror-mist transition is a probabilistic crossover, which is rather sharp due to the super-exponential variation of the localized branching probability for $G\!<\!G_{\rm B}$. Transiently-excited localized branching events in this regime are spatially uncorrelated and small, featuring a scale comparable to the disorder correlation length $R$, responsible for the misty appearance of the fracture surface. The mist-hackle transition occurs at $G\!\simeq\!G_{\rm B}$, upon which localized branching events become longer-lived and feature dynamic lengthscales that increase with increasing $G$. The wider and longer branching events are responsible for the hackle-like appearance of the fracture surface. Finally, at sufficiently large $G$, branches can capture the entire system thickness $L_z$, statistically recovering translational symmetry in the thickness direction. This 3D-to-2D transition results in macro-branching.

Our findings also raise some pressing questions for future investigation. A central element in the emerging physical picture is finite-amplitude quenched disorder. In the absence of well-established data about spatial disorder in the fracture energy of amorphous materials, we invoked in our calculations a minimal assumption in the form of a Gaussian disorder characterized by a dimensionless strength $\sigma$ and correlation length $R$. It would be most desirable to quantitatively characterize the statistical properties of various important physical quantities in amorphous materials, including the amplitude of fluctuations and their spatial correlation length. A possible indirect way to address this point might be to study surface roughness at low propagation velocities~\cite{bonamy.11,Ponson2023}.

Furthermore, we have shown that various quantities, e.g., the onset of localized branching velocity and the limiting velocity, are disorder dependent. Consequently, extracting the properties of the disorder may lead to a quantitative comparison with experiments. Finally, while our large-scale calculations have been carried out using cutting-edge GPU-based computational platforms~\cite{SI}, we cannot yet rule some finite-size effects in our results, especially with respect to $L_y$, but possibly also to $L_z$. Future research, potentially employing enhanced computational power, should shed light on this issue as well.\\

{\em Acknowledgements}. This work has been supported by the United States-Israel Binational Science Foundation (BSF, grant no.~2018603). E.B.~acknowledges support from the Ben May Center for Chemical Theory and Computation, and the Harold Perlman Family.

\clearpage

\onecolumngrid
\begin{center}
              \textbf{\Large Supplemental materials}
\end{center}

\setcounter{equation}{0}
\setcounter{figure}{0}
\setcounter{section}{0}
\setcounter{subsection}{0}
\setcounter{table}{0}
\setcounter{page}{1}
\makeatletter
\renewcommand{\theequation}{S\arabic{equation}}
\renewcommand{\thefigure}{S\arabic{figure}}
\renewcommand{\thesection}{S-\Roman{section}}
\renewcommand{\thesubsection}{S-\Roman{subsection}}
\renewcommand*{\thepage}{S\arabic{page}}
\twocolumngrid

The goal of this document is to provide some technical details regarding the results presented in the manuscript and to offer some additional supporting data.

\subsection{The 3D phase-field framework}
\label{sec:PF}

The 3D phase-field framework we employed has been very recently presented in great detail in the Supplementary Materials file of~\cite{das2023dynamics}. Here, for completeness, we very briefly repeat the main elements of the formulation and highlight specific features of its application in this work. A general material is described in this framework by the following potential energy $U$, kinetic energy $T$ and dissipation function $D$~\cite{chen2017,lubomirsky2018,vasudevan2021oscillatory}
\begin{eqnarray}
\label{Eq:Lagrangian_U}
U&=&\int \left[\frac{1}{2}\kappa\left(\nabla\phi\right)^{2}+ g(\phi)\,e({\bm u}) + w(\phi)\,e_{\rm c}\right]dV \ ,\quad \\
\label{Eq:Lagrangian_T}
T&=&\int\!\frac{1}{2}f(\phi)\,\rho\left(\partial_t {\bm u}\right)^2 dV \ ,\quad \\
\label{Eq:dissipation}
D&=&\frac{1}{2\chi}\int \left(\partial_t \phi\right)^{2}dV \ ,\quad
\end{eqnarray}
in terms of a 3D time-dependent vectorial displacement field $\bm{u}(x,y,z,t)$ and a 3D time-dependent auxiliary scalar phase-field $0\!\le\!\phi(x,y,z,t)\!\le\!1$. Here, $dV$ is a volume differential and the integration extends over the entire system. The evolution of $\phi({\bm x},t)$ and ${\bm u}({\bm x},t)$ follows Lagrange's equations
\begin{eqnarray}
\frac{\partial}{\partial t}\left[\frac{\delta L}{\delta\left(\partial\psi/\partial t\right)}\right]-\frac{\delta L}{\delta\psi}
+\frac{\delta D}{\delta\left(\partial\psi/\partial t\right)}=0 \ ,
\label{Eq:Lagrange_eqs}
\end{eqnarray}
where $L\=T-U$ is the Lagrangian and $\psi\=(\phi,u_x,u_y,u_z)$, i.e.~${\bm u}\=(u_x,u_y,u_z)$ are the components of the displacement vector field.

An intact/unbroken material state corresponds to $\phi\=1$, for which $g(\phi)\=f(1)\=1\!-\!w(1)\=1$. It describes a non-dissipative, elastodynamic material response characterized by an energy density $e({\bm u})$. For the latter, we use the linear elastic energy density
\begin{equation}
e({\bm u})=\frac{1}{2}\lambda\,\text{tr}^2({\bm \varepsilon}) + \mu\,\text{tr}({\bm \varepsilon}) \ ,
\label{eq:LE}
\end{equation}
where ${\bm \varepsilon}\=\tfrac{1}{2}[{\bm \nabla}{\bm u}+({\bm \nabla}{\bm u})^{\rm T}]$ is the infinitesimal (linearized) strain tensor, and $\lambda$ and $\mu$ (shear modulus) are the Lam\'e coefficients. We set $\lambda\=2\mu$ in all of our calculations.

Dissipation, loss of load-bearing capacity and eventually fracture accompanied by the generation of traction-free surfaces are associated with a strain energy density threshold $e_{\rm c}$. When the latter is surpassed, $\phi$ decreases from unity and the degradation functions $g(\phi)$, $f(\phi)$ and $1\!-\!w(\phi)$ also decrease from unity towards zero, upon which complete fracture takes place. We adopt the so-called KKL choice of the degradation functions~\cite{karma2001phase,vasudevan2021oscillatory}, corresponding to $f(\phi)\=g(\phi)$ and $w(\phi)\=1-g(\phi)$, with $g(\phi)\=4\phi^3-3\phi^4$. This choice, together with the linear elastic strain energy density in Eq.~\eqref{eq:LE} (that implies a vanishing nonlinear elastic zone near the crack front~\cite{chen2017,lubomirsky2018,vasudevan2021oscillatory}), suppress the 2D oscillatory instability~\cite{vasudevan2021oscillatory}.

The resulting set of nonlinear partial differential field equations feature a dissipation lengthscale $\xi\=\sqrt{\kappa/2e_{\rm c}}$ near crack fronts and an associated dissipation timescale $\tau\=(2\chi e_{\rm c})^{-1}$. Upon expressing length in units of $\xi$, time in units of $\xi/c_{\rm s}$, energy density in units of $\mu$ and the mass density $\rho$ in units of $\mu/c_{\rm s}^2$ ($c_{\rm s}\=\sqrt{\mu/\rho}$ is the shear wave-speed), the dimensionless set of equations depends on just two dimensionless parameters: $e_{\rm c}/\mu$ and $\beta\=\tau\,c_{\rm s}/\xi$. The latter controls the $v$-dependence of the fracture energy, $\Gamma(v)$~\cite{chen2017,lubomirsky2018,vasudevan2021oscillatory}. In our calculations, we set $e_{\rm c}/\mu\=0.01$ and $\beta\=2.76$. We also set $L_y\=256\xi$ and $L_z\=179\xi$, where $L_x$ is essentially indefinitely large due to an employed treadmill procedure~\cite{vasudevan2021oscillatory}. The boundary conditions are specified in the manuscript and the details of the numerical implementation are provided in~\cite{das2023dynamics}. Each calculation is perform on a single GPU (NVIDIA RTX A6000 or NVIDIA A40), either own by the Bouchbinder group or available on one of the computer clusters at Weizmenn Institute of Science. A typical simulation time (e.g., one of those shown on the leftmost column of Fig.~3a in the manuscript) is $\sim\!5$ days.

\subsection{Continuous quenched disorder}
\label{sec:disorder}

In~\cite{das2023dynamics}, we considered only isolated/discrete material heterogeneities (asperities). In the present work, we incorporated continuous quenched disorder into the phase-field framework, which was shown in the manuscript to play essential roles in 3D dynamic fracture. Continuous quenched disorder is introduced in two steps. First, as in~\cite{das2023dynamics}, we define a dimensionless auxiliary quenched disorder field $\zeta({\bm x})$, which can be coupled to any physical parameter in the problem. This coupling is achieved by transforming an originally spatially uniform parameter $\alpha_0$ into a field of the form $\alpha({\bm x})\=\alpha_0[1+\alpha_{_{\zeta}} \zeta({\bm x})]$, where $0\!\le\!\alpha_{_{\zeta}}\!\le\!1$ is a dimensionless coupling coefficient.

In the manuscript, $-1\!\le\!\zeta({\bm x})$ is taken to follow a Gaussian distribution of width $\sigma$ and zero mean, i.e., $\zeta$ at each spatial point ${\bm x}$ (in practice, at each numerical grid point) is extracted from a normal distribution with standard deviation $\sigma$ and zero mean, with no spatial correlations. That is, $\zeta$ at a given spatial location is independent of its value at other locations. The value of $\sigma$ is used to quantify the strength of disorder throughout the manuscript. Note that the choice $-1\!\le\!\zeta({\bm x})$ ensures that $\alpha({\bm x})\!\ge\!0$ for any $0\!\le\!\alpha_{_{\zeta}}\!\le\!1$, and hence the procedure can be naturally applied to positive physical quantities (e.g., the fracture energy and/or the shear modulus). The Gaussian probability associated with $-\infty\!<\!\zeta({\bm x})\!<\!-1$ is small in our calculations, and the way it is taken into account is further discussed below.

In the second step, spatial correlations in the quenched disorder are introduced. To this aim, we define a spatial kernel $K({\bm x})$ of compact support $R$, of the form $K({\bm x})\=1-(|{\bm x}|/R)^5$ for $|{\bm x}|\!\le\!R$ and zero otherwise. Then the actual value of the physical parameter of interest at a point ${\bm x}$, say $\bar{\alpha}({\bm x})$, is obtained through a convolution of $\alpha({\bm x})$ with the kernel $K({\bm x})$, i.e., $\bar{\alpha}({\bm x})\=K({\bm x})\ast \alpha({\bm x})$. Consequently, $\bar{\alpha}({\bm x})$ features a spatial correlation length $R$. In the manuscript, we set $\alpha_{_{\zeta}}\=0.9$ (that renormalizes the disorder strength) and $R\=10\xi$. The above procedure for introducing quenched disorder in any physical parameter is applied in the manuscript for the quasi-static ($v\!\to\!0$) fracture energy $\Gamma_0$.

\begin{figure}[h]
\center
\includegraphics[width=0.49\textwidth]{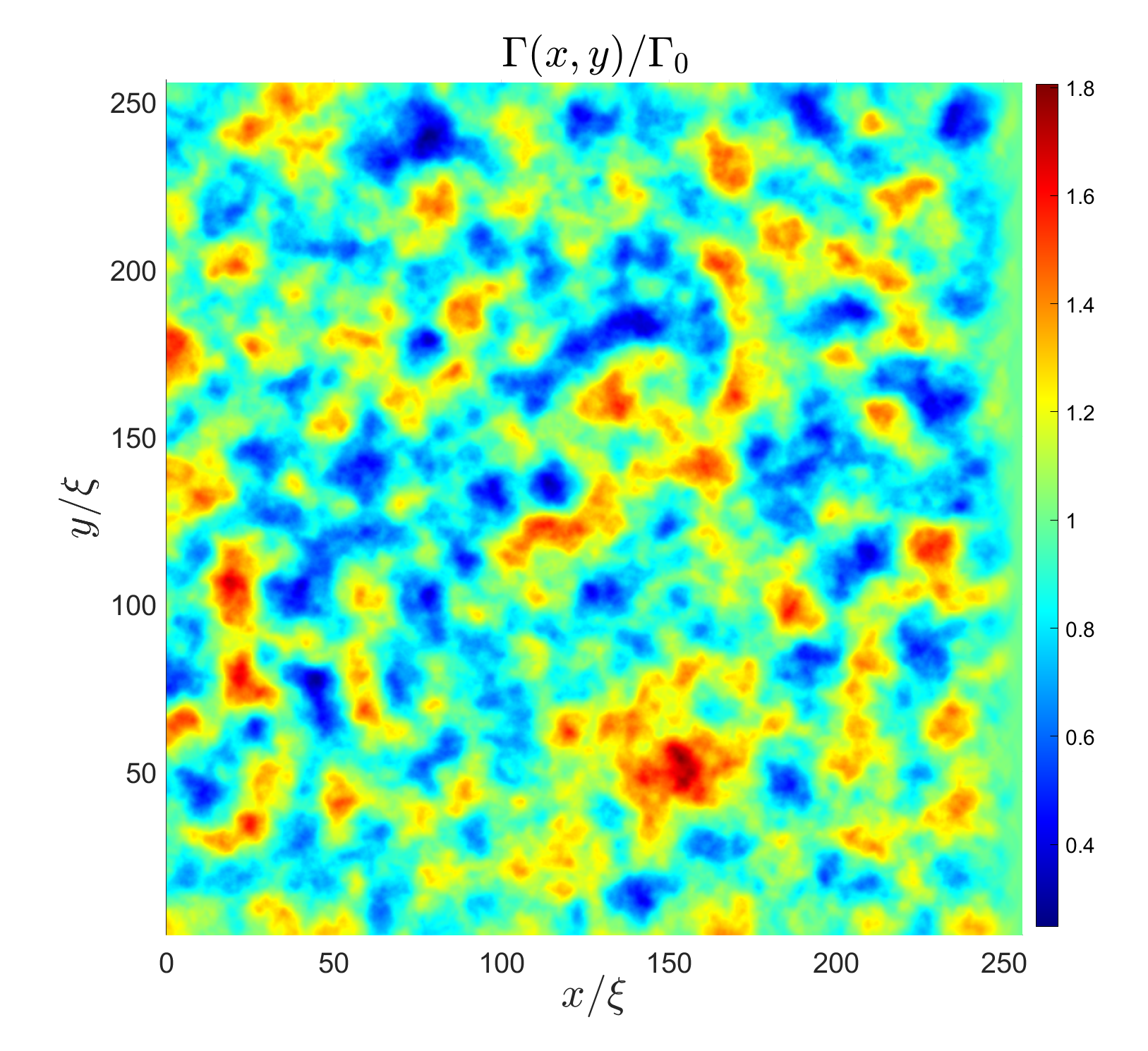}
\caption{A 2D $x\!-\!y$ cut of a single realization of the $\Gamma({\bm x})$ disorder field generated with $\sigma\!=\!0.25$ and $R\!=\!10\xi$. The spatial correlation length $R$ of the quenched disorder is evident.}
\label{fig:figS1}
\end{figure}

It is known that $\Gamma_0\!\sim\!e_{\rm c}\xi\!\sim\!\sqrt{\kappa e_{\rm c}}$~\cite{chen2017,lubomirsky2018,vasudevan2021oscillatory} and recall that $\beta\=\tau c_{\rm s}/\xi\!\sim\!(\chi e_{\rm c} \xi)^{-1}$ (i.e., with $\tau\!\sim\!(\chi e_{\rm c})^{-1}$). Consequently, we can simultaneously couple $\kappa$, $e_{\rm c}$ and $\chi$ to the very same realization of the quenched disorder field $\zeta({\bm x})$ (with the same $\alpha_{_{\zeta}}$) such that the quasi-static fracture energy features quenched disorder, while $\xi\!\sim\!\sqrt{\kappa/e_{\rm c}}$ and $\beta\!\sim\!(\chi e_{\rm c} \xi)^{-1}$ are kept fixed (i.e., independent of ${\bm x}$). This way, the dimensionless quasi-static fracture energy --- denoted by $\bar{\Gamma}/\Gamma_0$ --- follows a Gaussian distribution with unity mean, standard deviation $\alpha_{_{\zeta}}\sigma$ and spatial correlation length $R$, as stated in the manuscript, while its rate dependence (variation with $v$, which is controlled by $\beta$) is spatially uniform. A 2D cut of a single realization of the $\Gamma({\bm x})$ disorder field is presented in Fig.~\ref{fig:figS1}.

\subsection{The average crack velocity and the quantification of out-of-plane fracture structures}
\label{subsec:quantification}

\hspace{-0.3cm}{\bf A}. {\em The average crack velocity}

\begin{figure}[h]
\center
\includegraphics[width=0.49\textwidth]{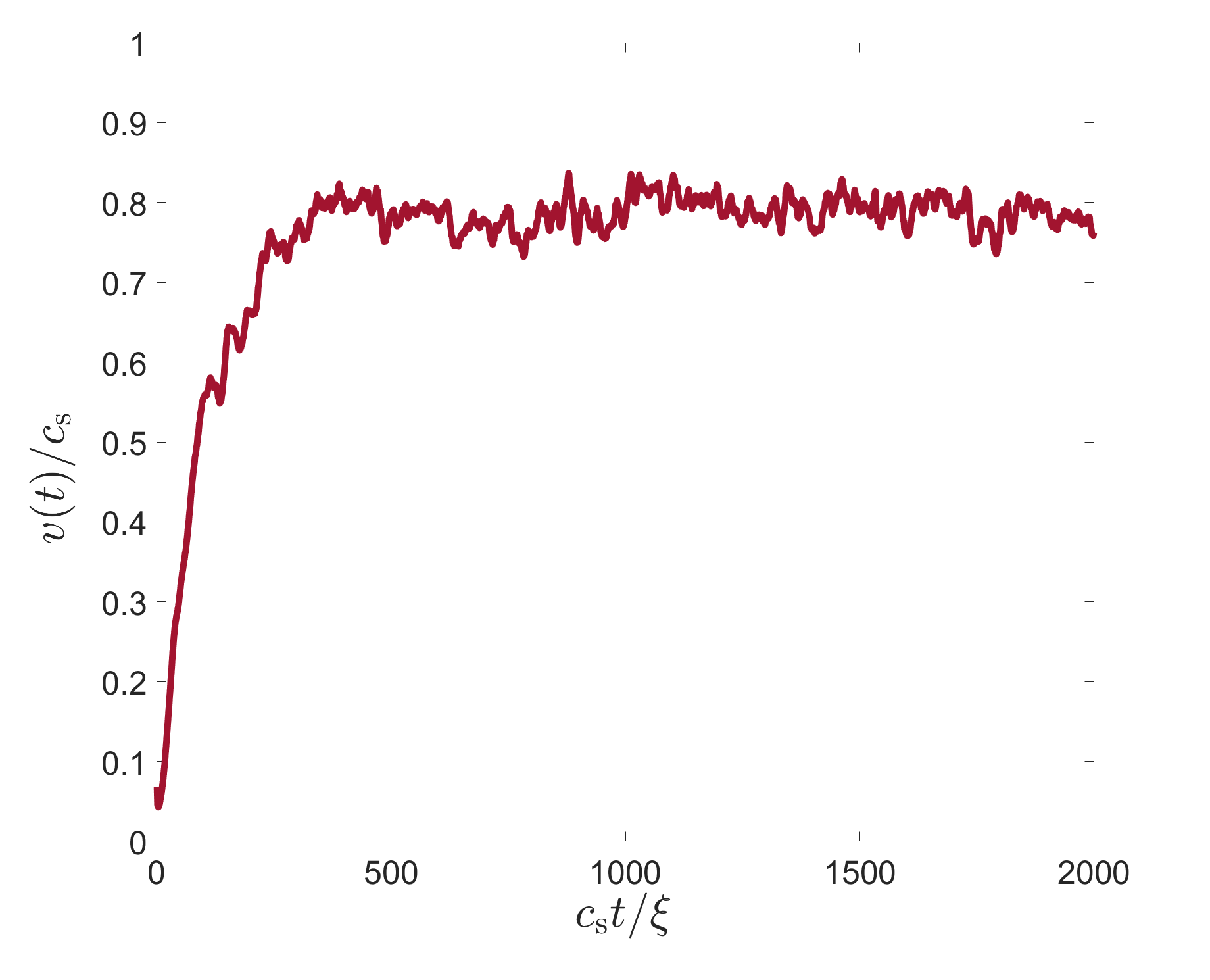}
\caption{The average crack front velocity $v(t)/c_{\rm s}$, as defined in the text, plotted against $c_{\rm s}t/\xi$ for $G/\Gamma_0\!=\!2.25$ (with our generic quenched disorder parameters $\sigma\!=\!0.25$ and $R\!=\!10\xi$). It is observed that after an initial acceleration phase, the crack settles into a statistical steady state (defining the mean velocity $v$ per $G$, accompanied by temporal fluctuations), where our analysis is performed.}
\label{fig:figS2}
\end{figure}

A central quantity in the manuscript is the average crack front velocity $v$. In the presence of quenched disorder, the crack front does not generally remain a continuous line propagating in a 3D space. As shown in the manuscript, not only it can be seriously distorted and meander out of the symmetry plane, but it actually undergoes various topological changes as the crack develops localized branches, hierarchial facets, segmentation and more. In view of these complex structures, the average crack front velocity $v$ is operationally defined as follows: at each point in time $t$, the crack is defined through the phase-field $\phi({\bm x},t)\!=\!1/2$ iso-surface. Then, we find the intersection of this iso-surface with the $x\!-\!y$ plane associated with every $z$ value. The intersection point with the largest $x$ coordinate (the crack propagates in the positive $x$ direction) corresponds to the front position ${\bm f}(z,t)\=(f_x(z,t),f_y(z,t))$. Finally, the average crack front velocity $v(t)$ is obtained as the $z$-average of $\partial_t f_x(z,t)$, i.e., $v(t)\=\langle\partial_t f_x(z,t)\rangle_z$. When the crack reaches a statistical steady state, $v(t)$ features a plateau that defines the steady-state velocity $v$ (for each $G$), accompanied by temporal fluctuations, as demonstrated in Fig.~\ref{fig:figS2}.

\hspace{-0.3cm}{\bf B}. {\em The quantification of out-of-plane fracture structures}

The uppermost out-of-plane crack structures shown in Fig.~3 in the manuscript are obtained as follows: the $\phi({\bm x})\!=\!1/2$ iso-surface at the end of each phase-field simulation is considered. Then, at each $(x,z)$ point we construct a vertical line along $y$ and find the uppermost $y$ intersection with the $\phi({\bm x})\!=\!1/2$ iso-surface. The resulting surface is plotted in Fig.~3 for different $G$ values. The apparent areal ratio, plotted in Fig.~5 in the manuscript (left $y$-axis), is obtained as the total area of the uppermost out-of-plane surface divided by the corresponding nominal (planar) area, as noted in the manuscript. The true areal ratio is in fact larger
as secondary structures develop beneath the uppermost out-of-plane crack surface at large $G$, as noted in the manuscript.

Finally, calculating the out-of-plane fluctuations --- that are plotted in Fig.~5 in the manuscript (right $y$-axis) --- involves also the lowermost out-of-plane surface, which is obtained similarly to the uppermost one, just for the lowermost $y$ intersection of vertical lines with the $\phi({\bm x})\!=\!1/2$ iso-surface (note that the uppermost and lowermost surfaces are not symmetric, i.e., the up-down symmetry is broken, as is further discussed below). The average over $(x,z)$ of the difference between the uppermost and lowermost surfaces (in units of $\xi$) provides a measure of out-of-plane fluctuations.

\subsection{A simple probabilistic model of localized branching in 3D}
\label{sec:model}

As explained above, the auxiliary quenched disorder field $\zeta({\bm x})$ is initially extracted from a Gaussian distribution of zero mean and standard deviation $\sigma$, $\frac{1}{\sqrt{2\pi}\sigma}\exp\!{(\!-\tfrac{1}{2}\zeta^{2}/\bar\sigma^{2})}$, independently for each spatial location ${\bm x}$. This Gaussian probability distribution function is normalized over the domain $-\infty\!<\!\zeta\!<\!\infty$. However, we defined for a physical parameter $\alpha$ the following disordered field $\alpha({\bm x})/\alpha_0\=1+\alpha_{_{\zeta}}\zeta({\bm x})$, with $0\!\le\!\alpha_{_{\zeta}}\!\le\!1$, which can become negative for sufficiently negative values of $\zeta$. As in our case $\alpha$ corresponds to the quasi-static fracture energy, which is a positive quantity, to ensure its positivity for any $\alpha_{_{\zeta}}$ we restrict the Gaussian distribution to be valid for $-1\!<\!\zeta\!<\!\infty$. To conform with the conservation of probability, we attribute the probability $\intop_{-\infty}^{-1}\frac{1}{\sqrt{2\pi}\sigma}\exp\!{(\!-\tfrac{1}{2}\zeta^{2}/\bar\sigma^{2})}\,d\zeta$ to $\zeta\=-1$. This probability (corresponding to a probability density proportional to a delta-function, $\delta(\zeta+1)$) constitutes a small addition to the Gaussian distribution over $-1\!<\!\zeta\!<\!\infty$ for the $\sigma$ values we considered in the manuscript. Finally, as explained above, the disordered field is convolved with a compact support kernel, endowing it with a spatial correlation length $R$.

This quenched disorder generation procedure is applied in the manuscript to the quasi-static fracture energy, i.e., $\alpha/\alpha_0$ is identified with $\bar{\Gamma}/\Gamma_0$. When the probability distribution $p(\bar{\Gamma}/\Gamma_0)$ in 3D is considered together with the 2D homogeneous-material branching instability, which occurs when the dimensionless crack driving force $G/\Gamma_0$ surpasses a threshold $G_{\rm B}/\Gamma_0$, a simple probabilistic model of localized branching in 3D can be constructed. The model consists of three ingredients: (i) Over a characteristic lengthscale comparable to the correlation length $R$, the 3D material effectively corresponds to a 2D homogeneous material. (ii) Following (i), fracture energy fluctuations of characteristic scale $R$ can lead to 3D localized branching $G/\bar\Gamma\!>\!G_{\rm B}/\Gamma_0$ at a given crack driving force $G$. (iii) Localized branching events in 3D are largely independent of each other for $G\!<\!G_{\rm B}$.

The probability distribution of $\bar{\Gamma}/\Gamma_0$ over the correlation length $R$ remains Gaussian, yet with a renormalized standard deviation $\bar\sigma(\sigma,R)\!<\!\sigma$, which one can calculate. Rearranging the inequality in (ii) to read $\bar\Gamma/\Gamma_0\!<\!G/G_{\rm B}$ and invoking (iii), we can use $p(\bar{\Gamma}/\Gamma_0)$ over the correlation length $R$ to estimate the localized branching instability in 3D, $p(G;\sigma)$, as
\begin{eqnarray}
p(G;\sigma)&\simeq&\!\!\!\!\!\intop_{-1}^{\frac{G/G_{\rm B}-1}{\alpha_{_{\!\zeta}}}}\frac{e^{-\tfrac{1}{2}\zeta^{2}/\bar\sigma^{2}}}{\sqrt{2\pi}\bar\sigma}\,d\zeta+
\intop_{-\infty}^{-1}\frac{e^{-\tfrac{1}{2}\zeta^{2}/\bar\sigma^{2}}}{\sqrt{2\pi}\bar\sigma}\,d\zeta  \\
&=&\intop_{-\infty}^{\frac{G/G_{\rm B}-1}{\alpha_{_{\!\zeta}}\bar\sigma}}\frac{e^{-\tfrac{1}{2}\eta^{2}}}{\sqrt{2\pi}}\,d\eta = \frac{1}{2}\left[1+\text{erf}\left(\frac{G/G_{\rm B}-1}{\sigma_{\!_{\rm R}}}\right)\right] \nonumber \ ,\\ \nonumber
\end{eqnarray}
with $\sigma_{\!_{\rm R}}\=\sqrt{2}\,\alpha_{_{\zeta}}\bar\sigma(\sigma,R)$, as in Eq.~(2) in the manuscript. For $\alpha_{_{\zeta}}\=0.9$, $\sigma\=0.25$ and $R\=10\xi$ used in the manuscript, we have $\sigma_{\!_{\rm R}}\!=\!0.233$, employed in Fig.~3a therein.

\subsection{Additional supporting results}
\label{sec:results}

In this section, we present additional results that further support those presented in the manuscript.

\vspace{0.3cm}
\hspace{-0.3cm}{\bf A}. {\em A fractographic signature of crack front waves}

Crack front waves (FWs) are spatiotemporal objects that propagate along moving crack fronts in 3D, featuring coupled in- and out-of-plane components~\cite{sharon2001propagating,sharon2002crack,fineberg2003crack,livne2005universality,das2023dynamics}. One process by which a pair of FWs is spontaneously generated is micro-branching events~\cite{fineberg2003crack,livne2005universality}, where the out-of-plane component of each FW leaves a fractographic signature on the fracture surface. It takes the form of two linear (straight) tracks emanating from the micro-branch and forming an opening angle $2\gamma$ between them, resulting in V-shaped tracks~\cite{fineberg2003crack,livne2005universality}. The FW velocity (in the laboratory frame of reference) $c_{_{\rm FW}}$ is related to the crack propagation velocity $v$ according to $c_{_{\rm FW}}\=v/\!\cos(\gamma)$, where $c_{_{\rm FW}}$ was found to be close to the Rayleigh wave-speed $c_{_{\rm R}}$, very weakly dependent on $v$~\cite{ramanathan.97,morrissey1998,adda-bedia.13,das2023dynamics}.

\begin{figure}[h]
\center
\includegraphics[width=0.48\textwidth]{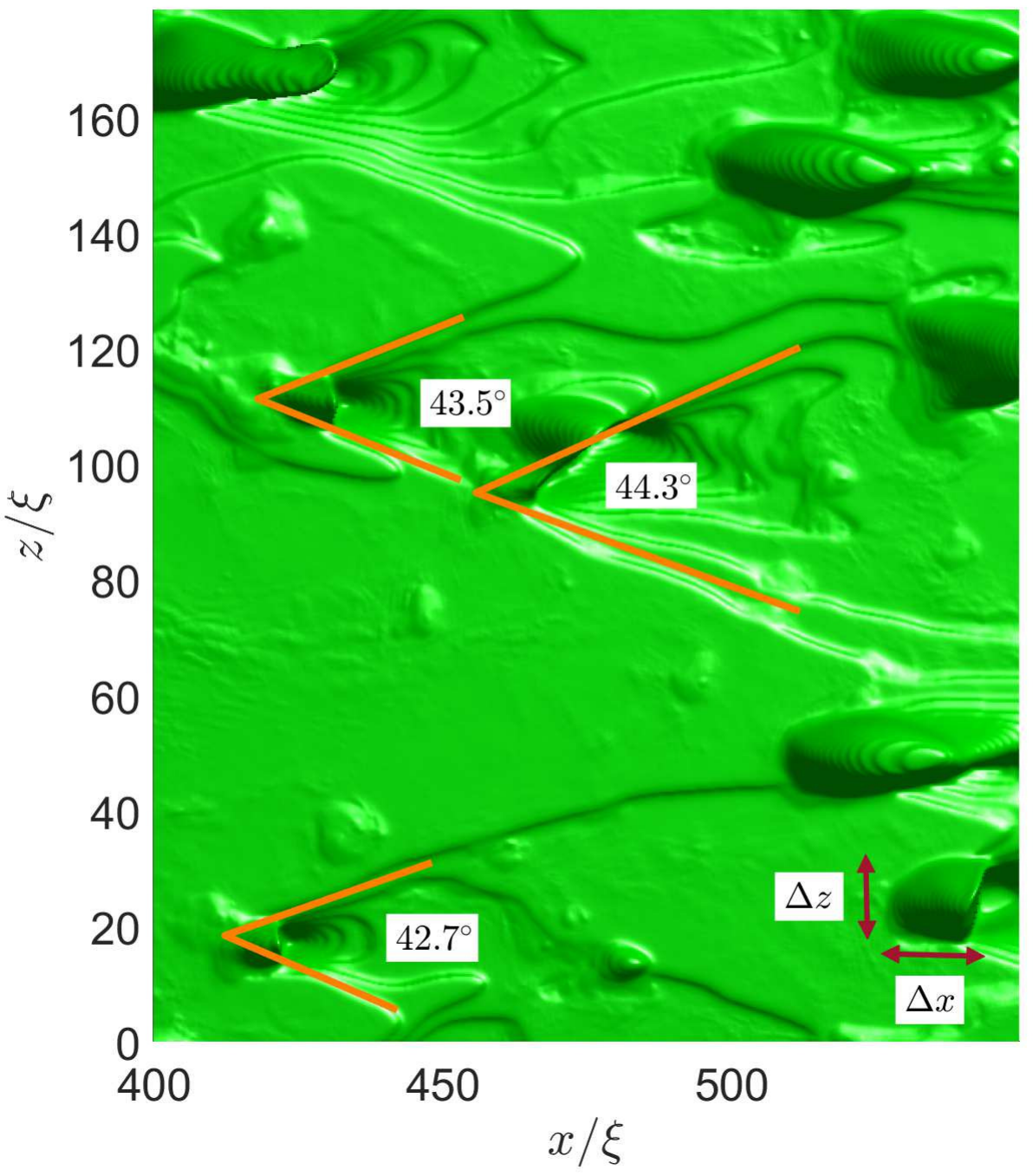}
\caption{A zoom in on a few localized branching events originally presented in Fig.~3b and Fig.~3f in the manuscript (top view). For three localized branching events, the V-shaped tracks (see text for details) are marked and their opening angle $2\gamma$ is indicated. The latter, corresponding to $\gamma\!\simeq\!21.75^\circ$, is essentially the same for the three events. The localized branch width $\Delta{z}$ and length $\Delta{x}$ are illustrated for the localized branching event in the bottom-right corner. Results involving $\Delta{z}$ and $\Delta{x}$ are presented in Fig.~4b and the inset of Fig.~5 in the manuscript, and discussed therein.}
\label{fig:figS3}
\end{figure}

In Fig.~3b in the manuscript, it was observed that isolated localized branching events in our simulations are accompanied by V-shaped tracks on the fracture surface. Our goal here is to test whether these tracks are consistent with experimental observations regarding FWs. To that aim, we present in Fig.~\ref{fig:figS3} a zoom in on a few localized branching events originally shown in Fig.~3b and Fig.~3f in the manuscript, where the V-shaped tracks with an opening angle $2\gamma$ are clearly observed. We find that $\gamma\!\simeq\!21.75^\circ$, which is essentially the same for all localized branching events analyzed (see Fig.~\ref{fig:figS3}). The instantaneous (not the mean) crack propagation velocity at this point in the dynamics was $v\=0.835\pm0.015 c_{\rm s}$, such that we find $c_{_{\rm FW}}\=v/\!\cos(\gamma)\=0.9c_{\rm s}\!\simeq\!0.97c_{_{\rm R}}$, which is indeed consistent with the experimentally observed velocity of FWs.

\vspace{0.3cm}
\hspace{-0.3cm}{\bf B}. {\em Asymmetric localized branch profiles}

Some experimental evidence indicates that micro-branches are asymmetric (with respect to the main crack plane), at least at not too high crack propagation velocities, e.g., see the middle panel (center) in Fig.~4c in~\cite{sharon1996microbranching}. That is, micro-branches in this regime do not appear to be formed in pairs that propagate predominantly symmetrically, but rather single micro-branches tend to form and either propagate upwards or downwards relative to the main crack plane. It has been experimentally shown that the micro-branches approximately follow a $y\!\sim\!x^{0.7}$ profile out of the crack plane ($y\=0$, where $x$ is the propagation direction).

\begin{figure}[h]
\center
\includegraphics[width=0.49\textwidth]{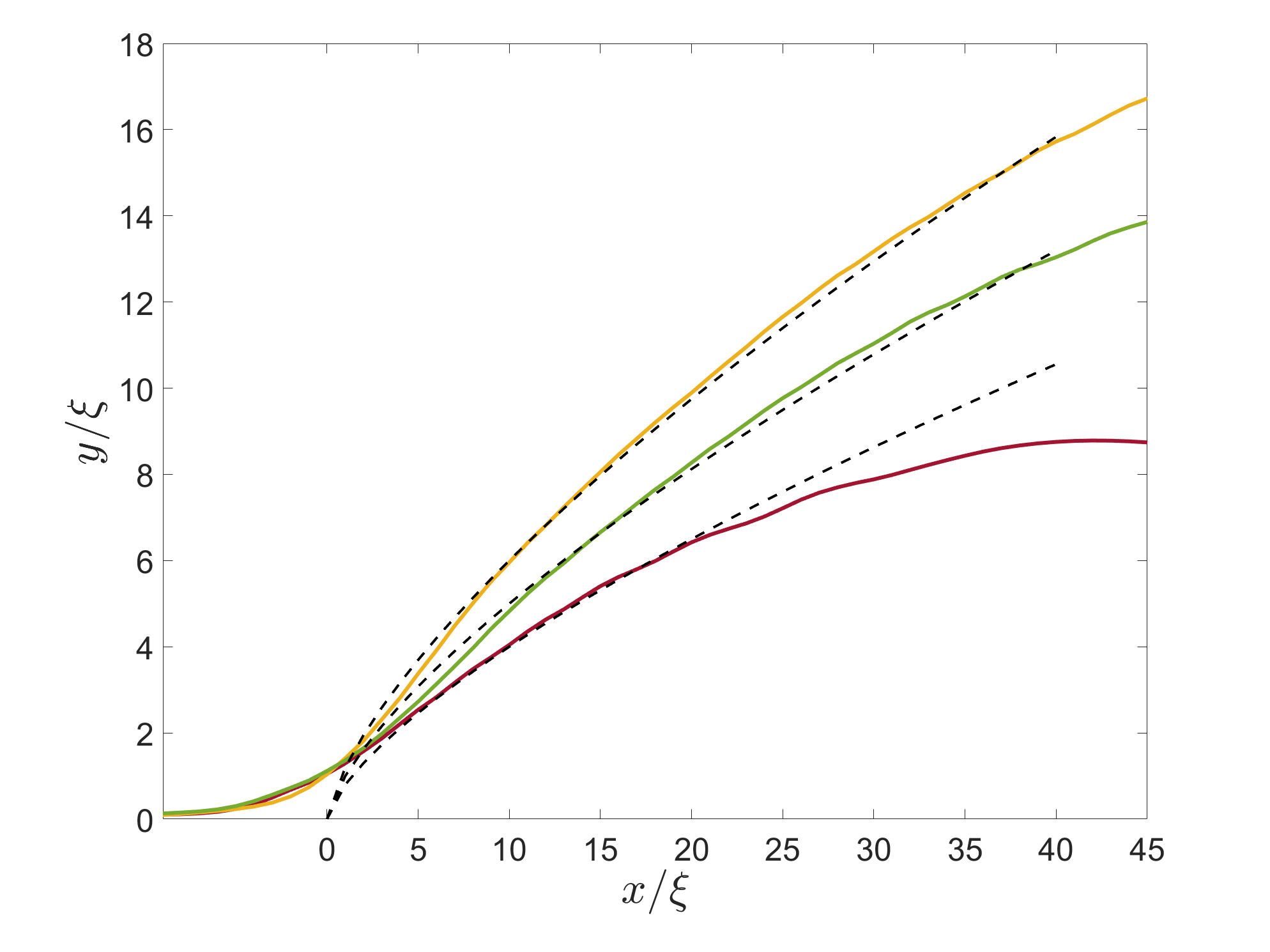}
\caption{Three examples of the $x\!-\!y$ profile of asymmetric localized branches (in the $z$ direction) randomly selected from our simulations (and shifted such that they all overlap at their early stages of development). The dashed lines are guides to eye corresponding to $y\!\sim\!x^{0.7}$, as found experimentally (e.g., see Fig.~12 in~\cite{sharon1996microbranching} and Fig.~2 in~\cite{sharon1998universal}), each with a different amplitude.}
\label{fig:figS4}
\end{figure}

Many of the localized branching events in our simulations, especially in the not too high velocities regime, are indeed asymmetric. That is, even if a pair of localized branches initially form, the quenched disorder and/or the interaction with other out-of-plane structures break the up-down symmetry, resulting in a single dominant localized branch. In these situations, we expect the approximate $y\!\sim\!x^{0.7}$ profile --- that emerges from the elastodynamic interaction of the localized branch with the main crack --- to be observed in our case as well. This is indeed demonstrated in Fig.~\ref{fig:figS4}


\begin{thebibliography}{10}
\expandafter\ifx\csname url\endcsname\relax
  \def\url#1{\texttt{#1}}\fi
\expandafter\ifx\csname urlprefix\endcsname\relax\def\urlprefix{URL }\fi
\providecommand{\bibinfo}[2]{#2}
\providecommand{\eprint}[2][]{\url{#2}}

\bibitem{ravi1998dynamic}
\bibinfo{author}{Ravi-Chandar, K.}
\newblock \bibinfo{title}{Dynamic fracture of nominally brittle materials}.
\newblock \emph{\bibinfo{journal}{International Journal of Fracture}}
  \textbf{\bibinfo{volume}{90}}, \bibinfo{pages}{83--102}
  (\bibinfo{year}{1998}).

\bibitem{fineberg.99}
\bibinfo{author}{Fineberg, J.} \& \bibinfo{author}{Marder, M.}
\newblock \bibinfo{title}{Instability in dynamic fracture}.
\newblock \emph{\bibinfo{journal}{Physics Reports}}
  \textbf{\bibinfo{volume}{313}}, \bibinfo{pages}{1--108}
  (\bibinfo{year}{1999}).

\bibitem{bonamy.11}
\bibinfo{author}{Bonamy, D.} \& \bibinfo{author}{Bouchaud, E.}
\newblock \bibinfo{title}{{Failure of heterogeneous materials: A dynamic phase
  transition?}}
\newblock \emph{\bibinfo{journal}{{Physics Reports}}}
  \textbf{\bibinfo{volume}{{498}}}, \bibinfo{pages}{{1--44}}
  (\bibinfo{year}{{2011}}).

\bibitem{bouchbinder.14}
\bibinfo{author}{Bouchbinder, E.}, \bibinfo{author}{Goldman, T.} \&
  \bibinfo{author}{Fineberg, J.}
\newblock \bibinfo{title}{The dynamics of rapid fracture: instabilities,
  nonlinearities and length scales}.
\newblock \emph{\bibinfo{journal}{Reports on Progress in Physics}}
  \textbf{\bibinfo{volume}{77}}, \bibinfo{pages}{046501}
  (\bibinfo{year}{2014}).

\bibitem{fineberg2015recent}
\bibinfo{author}{Fineberg, J.} \& \bibinfo{author}{Bouchbinder, E.}
\newblock \bibinfo{title}{Recent developments in dynamic fracture: Some
  perspectives}.
\newblock \emph{\bibinfo{journal}{International Journal of Fracture}}
  \textbf{\bibinfo{volume}{196}}, \bibinfo{pages}{33--57}
  (\bibinfo{year}{2015}).

\bibitem{freund}
\bibinfo{author}{Freund, L.~B.}
\newblock \emph{\bibinfo{title}{Dynamic Fracture Mechanics}}
  (\bibinfo{publisher}{Cambridge University Press},
  \bibinfo{address}{Cambridge}, \bibinfo{year}{1990}).

\bibitem{99bro}
\bibinfo{author}{Broberg, K.~R.}
\newblock \emph{\bibinfo{title}{Cracks and Fracture}}
  (\bibinfo{publisher}{Academic Press}, \bibinfo{address}{New York},
  \bibinfo{year}{1999}).

\bibitem{livne.07}
\bibinfo{author}{Livne, A.}, \bibinfo{author}{Ben-David, O.} \&
  \bibinfo{author}{Fineberg, J.}
\newblock \bibinfo{title}{Oscillations in rapid fracture}.
\newblock \emph{\bibinfo{journal}{Physical Review Letters}}
  \textbf{\bibinfo{volume}{98}}, \bibinfo{pages}{124301}
  (\bibinfo{year}{2007}).

\bibitem{lubomirsky2018}
\bibinfo{author}{Lubomirsky, Y.}, \bibinfo{author}{Chen, C.-H.},
  \bibinfo{author}{Karma, A.} \& \bibinfo{author}{Bouchbinder, E.}
\newblock \bibinfo{title}{Universality and stability phase diagram of
  two-dimensional brittle fracture}.
\newblock \emph{\bibinfo{journal}{Physical Review Letters}}
  \textbf{\bibinfo{volume}{121}}, \bibinfo{pages}{134301}
  (\bibinfo{year}{2018}).

\bibitem{bouchbinder2008}
\bibinfo{author}{Bouchbinder, E.}, \bibinfo{author}{Livne, A.} \&
  \bibinfo{author}{Fineberg, J.}
\newblock \bibinfo{title}{Weakly nonlinear theory of dynamic fracture}.
\newblock \emph{\bibinfo{journal}{Physical Review Letters}}
  \textbf{\bibinfo{volume}{101}}, \bibinfo{pages}{264302}
  (\bibinfo{year}{2008}).

\bibitem{bouchbinder.09b}
\bibinfo{author}{Bouchbinder, E.}
\newblock \bibinfo{title}{Dynamic crack tip equation of motion: High-speed
  oscillatory instability}.
\newblock \emph{\bibinfo{journal}{Physical Review Letters}}
  \textbf{\bibinfo{volume}{103}}, \bibinfo{pages}{164301}
  (\bibinfo{year}{2009}).

\bibitem{chen2017}
\bibinfo{author}{Chen, C.-H.}, \bibinfo{author}{Bouchbinder, E.} \&
  \bibinfo{author}{Karma, A.}
\newblock \bibinfo{title}{Instability in dynamic fracture and the failure of
  the classical theory of cracks}.
\newblock \emph{\bibinfo{journal}{Nature Physics}}
  \textbf{\bibinfo{volume}{13}}, \bibinfo{pages}{1186--1190}
  (\bibinfo{year}{2017}).

\bibitem{vasudevan2021oscillatory}
\bibinfo{author}{Vasudevan, A.}, \bibinfo{author}{Lubomirsky, Y.},
  \bibinfo{author}{Chen, C.-H.}, \bibinfo{author}{Bouchbinder, E.} \&
  \bibinfo{author}{Karma, A.}
\newblock \bibinfo{title}{Oscillatory and tip-splitting instabilities in
  \uppercase{2D} dynamic fracture: The roles of intrinsic material length and
  time scales}.
\newblock \emph{\bibinfo{journal}{Journal of the Mechanics and Physics of
  Solids}} \textbf{\bibinfo{volume}{151}}, \bibinfo{pages}{104372}
  (\bibinfo{year}{2021}).

\bibitem{ravi1984experimental_II}
\bibinfo{author}{Ravi-Chandar, K.} \& \bibinfo{author}{Knauss, W.}
\newblock \bibinfo{title}{An experimental investigation into dynamic fracture:
  \uppercase{II. M}icrostructural aspects}.
\newblock \emph{\bibinfo{journal}{International Journal of fracture}}
  \textbf{\bibinfo{volume}{26}}, \bibinfo{pages}{65--80}
  (\bibinfo{year}{1984}).

\bibitem{ravi1984experimental_III}
\bibinfo{author}{Ravi-Chandar, K.} \& \bibinfo{author}{Knauss, W.~G.}
\newblock \bibinfo{title}{An experimental investigation into dynamic fracture:
  \uppercase{III. O}n steady-state crack propagation and crack branching}.
\newblock \emph{\bibinfo{journal}{International Journal of fracture}}
  \textbf{\bibinfo{volume}{26}}, \bibinfo{pages}{141--154}
  (\bibinfo{year}{1984}).

\bibitem{scheibert2010brittle}
\bibinfo{author}{Scheibert, J.}, \bibinfo{author}{Guerra, C.},
  \bibinfo{author}{C{\'e}lari{\'e}, F.}, \bibinfo{author}{Dalmas, D.} \&
  \bibinfo{author}{Bonamy, D.}
\newblock \bibinfo{title}{Brittle-quasibrittle transition in dynamic fracture:
  An energetic signature}.
\newblock \emph{\bibinfo{journal}{Physical Review Letters}}
  \textbf{\bibinfo{volume}{104}}, \bibinfo{pages}{045501}
  (\bibinfo{year}{2010}).

\bibitem{guerra2012understanding}
\bibinfo{author}{Guerra, C.}, \bibinfo{author}{Scheibert, J.},
  \bibinfo{author}{Bonamy, D.} \& \bibinfo{author}{Dalmas, D.}
\newblock \bibinfo{title}{Understanding fast macroscale fracture from
  microcrack post mortem patterns}.
\newblock \emph{\bibinfo{journal}{Proceedings of the National Academy of
  Sciences}} \textbf{\bibinfo{volume}{109}}, \bibinfo{pages}{390--394}
  (\bibinfo{year}{2012}).

\bibitem{tanaka1998discontinuous}
\bibinfo{author}{Tanaka, Y.}, \bibinfo{author}{Fukao, K.},
  \bibinfo{author}{Miyamoto, Y.} \& \bibinfo{author}{Sekimoto, K.}
\newblock \bibinfo{title}{Discontinuous crack fronts of three-dimensional
  fractures}.
\newblock \emph{\bibinfo{journal}{EPL (Europhysics Letters)}}
  \textbf{\bibinfo{volume}{43}}, \bibinfo{pages}{664} (\bibinfo{year}{1998}).

\bibitem{steps2017}
\bibinfo{author}{Kolvin, I.}, \bibinfo{author}{Cohen, G.} \&
  \bibinfo{author}{Fineberg, J.}
\newblock \bibinfo{title}{Topological defects govern crack front motion and
  facet formation on broken surfaces}.
\newblock \emph{\bibinfo{journal}{Nature Materials}}
  \textbf{\bibinfo{volume}{17}}, \bibinfo{pages}{140--144}
  (\bibinfo{year}{2018}).

\bibitem{baumberger.08}
\bibinfo{author}{Baumberger, T.}, \bibinfo{author}{Caroli, C.},
  \bibinfo{author}{Martina, D.} \& \bibinfo{author}{Ronsin, O.}
\newblock \bibinfo{title}{Magic angles and cross-hatching instability in
  hydrogel fracture}.
\newblock \emph{\bibinfo{journal}{Physical Review Letters}}
  \textbf{\bibinfo{volume}{100}}, \bibinfo{pages}{178303}
  (\bibinfo{year}{2008}).

\bibitem{sharon1996microbranching}
\bibinfo{author}{Sharon, E.} \& \bibinfo{author}{Fineberg, J.}
\newblock \bibinfo{title}{Microbranching instability and the dynamic fracture
  of brittle materials}.
\newblock \emph{\bibinfo{journal}{Physical Review B}}
  \textbf{\bibinfo{volume}{54}}, \bibinfo{pages}{7128} (\bibinfo{year}{1996}).

\bibitem{sharon1998universal}
\bibinfo{author}{Sharon, E.} \& \bibinfo{author}{Fineberg, J.}
\newblock \bibinfo{title}{Universal features of the microbranching instability
  in dynamic fracture}.
\newblock \emph{\bibinfo{journal}{Philosophical Magazine B}}
  \textbf{\bibinfo{volume}{78}}, \bibinfo{pages}{243--251}
  (\bibinfo{year}{1998}).

\bibitem{sharon1999dynamics}
\bibinfo{author}{Sharon, E.} \& \bibinfo{author}{Fineberg, J.}
\newblock \bibinfo{title}{The dynamics of fast fracture}.
\newblock \emph{\bibinfo{journal}{Advanced Engineering Materials}}
  \textbf{\bibinfo{volume}{1}}, \bibinfo{pages}{119--122}
  (\bibinfo{year}{1999}).

\bibitem{livne2005universality}
\bibinfo{author}{Livne, A.}, \bibinfo{author}{Cohen, G.} \&
  \bibinfo{author}{Fineberg, J.}
\newblock \bibinfo{title}{Universality and hysteretic dynamics in rapid
  fracture}.
\newblock \emph{\bibinfo{journal}{Physical Review Letters}}
  \textbf{\bibinfo{volume}{94}}, \bibinfo{pages}{224301}
  (\bibinfo{year}{2005}).

\bibitem{lawn}
\bibinfo{author}{Lawn, B.}
\newblock \emph{\bibinfo{title}{Fracture of brittle solids}}
  (\bibinfo{publisher}{Cambridge University Press}, \bibinfo{year}{1993}).

\bibitem{johnson1968microstructure}
\bibinfo{author}{Johnson, J.} \& \bibinfo{author}{Holloway, D.}
\newblock \bibinfo{title}{Microstructure of the mist zone on glass fracture
  surfaces}.
\newblock \emph{\bibinfo{journal}{The Philosophical Magazine: A Journal of
  Theoretical Experimental and Applied Physics}} \textbf{\bibinfo{volume}{17}},
  \bibinfo{pages}{899--910} (\bibinfo{year}{1968}).

\bibitem{rabinovitch2000origin}
\bibinfo{author}{Rabinovitch, A.}, \bibinfo{author}{Belizovsky, G.} \&
  \bibinfo{author}{Bahat, D.}
\newblock \bibinfo{title}{Origin of mist and hackle patterns in brittle
  fracture}.
\newblock \emph{\bibinfo{journal}{Physical Review B}}
  \textbf{\bibinfo{volume}{61}}, \bibinfo{pages}{14968} (\bibinfo{year}{2000}).

\bibitem{jiao2015macroscopic}
\bibinfo{author}{Jiao, D.}, \bibinfo{author}{Qu, R.~T.} \&
  \bibinfo{author}{Zhang, Z.~F.}
\newblock \bibinfo{title}{Macroscopic bifurcation and fracture mechanism of
  polymethyl methacrylate}.
\newblock \emph{\bibinfo{journal}{Advanced Engineering Materials}}
  \textbf{\bibinfo{volume}{17}}, \bibinfo{pages}{1454--1464}
  (\bibinfo{year}{2015}).

\bibitem{SI}
 \bibinfo{title}{See Supplemental Materials (appearing in this PDF) for details}.

\bibitem{karma2001phase}
\bibinfo{author}{Karma, A.}, \bibinfo{author}{Kessler, D.} \&
  \bibinfo{author}{Levine, H.}
\newblock \bibinfo{title}{Phase-field model of mode \uppercase{III} dynamic
  fracture}.
\newblock \emph{\bibinfo{journal}{Physical Review Letters}}
  \textbf{\bibinfo{volume}{87}}, \bibinfo{pages}{45501} (\bibinfo{year}{2001}).

\bibitem{Karma2004}
\bibinfo{author}{Karma, A.} \& \bibinfo{author}{Lobkovsky, A.~E.}
\newblock \bibinfo{title}{Unsteady crack motion and branching in a phase-field
  model of brittle fracture}.
\newblock \emph{\bibinfo{journal}{Physical Review Letters}}
  \textbf{\bibinfo{volume}{92}}, \bibinfo{pages}{245510}
  (\bibinfo{year}{2004}).

\bibitem{Hakim.09}
\bibinfo{author}{Hakim, V.} \& \bibinfo{author}{Karma, A.}
\newblock \bibinfo{title}{Laws of crack motion and phase-field models of
  fracture}.
\newblock \emph{\bibinfo{journal}{Journal of the Mechanics and Physics of
  Solids}} \textbf{\bibinfo{volume}{57}}, \bibinfo{pages}{342--368}
  (\bibinfo{year}{2009}).

\bibitem{das2023dynamics}
\bibinfo{author}{Das, S.}, \bibinfo{author}{Lubomirsky, Y.} \&
  \bibinfo{author}{Bouchbinder, E.}
\newblock \bibinfo{title}{Dynamics of crack front waves in three-dimensional
  material failure}.
\newblock \emph{\bibinfo{journal}{Physical Review E}}
  \textbf{\bibinfo{volume}{108}}, \bibinfo{pages}{L043002}
  (\bibinfo{year}{2023}).

\bibitem{bleyer2017microbranching}
\bibinfo{author}{Bleyer, J.} \& \bibinfo{author}{Molinari, J.-F.}
\newblock \bibinfo{title}{Microbranching instability in phase-field modelling
  of dynamic brittle fracture}.
\newblock \emph{\bibinfo{journal}{Applied Physics Letters}}
  \textbf{\bibinfo{volume}{110}}, \bibinfo{pages}{151903}
  (\bibinfo{year}{2017}).

\bibitem{henry2013fractographic}
\bibinfo{author}{Henry, H.} \& \bibinfo{author}{Adda-Bedia, M.}
\newblock \bibinfo{title}{Fractographic aspects of crack branching instability
  using a phase-field model}.
\newblock \emph{\bibinfo{journal}{Physical Review E}}
  \textbf{\bibinfo{volume}{88}}, \bibinfo{pages}{060401}
  (\bibinfo{year}{2013}).

\bibitem{Ponson2023}
\bibinfo{author}{Ponson, L.}
\newblock \emph{\bibinfo{title}{Fracture Mechanics of Heterogeneous Materials:
  Effective Toughness and Fluctuations}}, \bibinfo{pages}{pp.~207--254 in
  Mechanics and Physics of Fracture: Multiscale Modeling of the Failure
  Behavior of Solids} (\bibinfo{publisher}{Springer International Publishing},
  \bibinfo{address}{Cham}, \bibinfo{year}{2023}).

\bibitem{roch2023dynamic}
\bibinfo{author}{Roch, T.}, \bibinfo{author}{Lebihain, M.} \&
  \bibinfo{author}{Molinari, J.-F.}
\newblock \bibinfo{title}{Dynamic crack-front deformations in cohesive
  materials}.
\newblock \emph{\bibinfo{journal}{Physical Review Letters}}
  \textbf{\bibinfo{volume}{131}}, \bibinfo{pages}{096101}
  (\bibinfo{year}{2023}).

\bibitem{lebihain2020effective}
\bibinfo{author}{Lebihain, M.}, \bibinfo{author}{Leblond, J.-B.} \&
  \bibinfo{author}{Ponson, L.}
\newblock \bibinfo{title}{Effective toughness of periodic heterogeneous
  materials: the effect of out-of-plane excursions of cracks}.
\newblock \emph{\bibinfo{journal}{Journal of the Mechanics and Physics of
  Solids}} \textbf{\bibinfo{volume}{137}}, \bibinfo{pages}{103876}
  (\bibinfo{year}{2020}).

\bibitem{Eshelby1970}
\bibinfo{author}{Eshelby, J.}
\newblock \emph{\bibinfo{title}{Energy relations and the energy-momentum tensor
  in continuum mechanics}}, \bibinfo{pages}{pp.~77--115 in Inelastic behaviour
  of solids} (\bibinfo{publisher}{McGraw-Hill}, \bibinfo{address}{New York},
  \bibinfo{year}{1970}).

\bibitem{adda-bedia.2007}
\bibinfo{author}{Katzav, E.}, \bibinfo{author}{Adda-Bedia, M.} \&
  \bibinfo{author}{Arias, R.}
\newblock \bibinfo{title}{Theory of dynamic crack branching in brittle
  materials}.
\newblock \emph{\bibinfo{journal}{International Journal of Fracture}}
  \textbf{\bibinfo{volume}{143}}, \bibinfo{pages}{245--271}
  (\bibinfo{year}{2007}).

\bibitem{kobayashi1974crack}
\bibinfo{author}{Kobayashi, A.}, \bibinfo{author}{Wade, B.},
  \bibinfo{author}{Bradley, W.} \& \bibinfo{author}{Chiu, S.}
\newblock \bibinfo{title}{Crack branching in homalite-100 sheets}.
\newblock \emph{\bibinfo{journal}{Engineering Fracture Mechanics}}
  \textbf{\bibinfo{volume}{6}}, \bibinfo{pages}{81--92} (\bibinfo{year}{1974}).

\bibitem{sun2021state}
\bibinfo{author}{Sun, Y.}, \bibinfo{author}{Edwards, M.~G.},
  \bibinfo{author}{Chen, B.} \& \bibinfo{author}{Li, C.}
\newblock \bibinfo{title}{A state-of-the-art review of crack branching}.
\newblock \emph{\bibinfo{journal}{Engineering Fracture Mechanics}}
  \textbf{\bibinfo{volume}{257}}, \bibinfo{pages}{108036}
  (\bibinfo{year}{2021}).

\bibitem{yoon2022situ}
\bibinfo{author}{Yoon, J.} \emph{et~al.}
\newblock \bibinfo{title}{In situ tensile and fracture behavior of monolithic
  ultra-thin amorphous carbon in tem}.
\newblock \emph{\bibinfo{journal}{Carbon}} \textbf{\bibinfo{volume}{196}},
  \bibinfo{pages}{236--242} (\bibinfo{year}{2022}).

\bibitem{wagner2011local}
\bibinfo{author}{Wagner, H.} \emph{et~al.}
\newblock \bibinfo{title}{Local elastic properties of a metallic glass}.
\newblock \emph{\bibinfo{journal}{Nature Materials}}
  \textbf{\bibinfo{volume}{10}}, \bibinfo{pages}{439--442}
  (\bibinfo{year}{2011}).

\bibitem{kapteijns2021elastic}
\bibinfo{author}{Kapteijns, G.}, \bibinfo{author}{Richard, D.},
  \bibinfo{author}{Bouchbinder, E.} \& \bibinfo{author}{Lerner, E.}
\newblock \bibinfo{title}{Elastic moduli fluctuations predict wave attenuation
  rates in glasses}.
\newblock \emph{\bibinfo{journal}{The Journal of Chemical Physics}}
  \textbf{\bibinfo{volume}{154}}, \bibinfo{pages}{081101}
  (\bibinfo{year}{2021}).

\bibitem{sharon2002crack}
\bibinfo{author}{Sharon, E.}, \bibinfo{author}{Cohen, G.} \&
  \bibinfo{author}{Fineberg, J.}
\newblock \bibinfo{title}{Crack front waves and the dynamics of a rapidly
  moving crack}.
\newblock \emph{\bibinfo{journal}{Physical Review Letters}}
  \textbf{\bibinfo{volume}{88}}, \bibinfo{pages}{085503}
  (\bibinfo{year}{2002}).

\bibitem{fineberg2003crack}
\bibinfo{author}{Fineberg, J.}, \bibinfo{author}{Sharon, E.} \&
  \bibinfo{author}{Cohen, G.}
\newblock \bibinfo{title}{Crack front waves in dynamic fracture}.
\newblock \emph{\bibinfo{journal}{International Journal of Fracture}}
  \textbf{\bibinfo{volume}{121}}, \bibinfo{pages}{55--69}
  (\bibinfo{year}{2003}).

\bibitem{adda-bedia.13}
\bibinfo{author}{Adda-Bedia, M.}, \bibinfo{author}{Arias, R.~E.},
  \bibinfo{author}{Bouchbinder, E.} \& \bibinfo{author}{Katzav, E.}
\newblock \bibinfo{title}{Dynamic stability of crack fronts: Out-of-plane
  corrugations}.
\newblock \emph{\bibinfo{journal}{Physical Review Letters}}
  \textbf{\bibinfo{volume}{110}}, \bibinfo{pages}{014302}
  (\bibinfo{year}{2013}).

\bibitem{boue2015origin}
\bibinfo{author}{Goldman~Bou{\'e}, T.}, \bibinfo{author}{Cohen, G.} \&
  \bibinfo{author}{Fineberg, J.}
\newblock \bibinfo{title}{Origin of the microbranching instability in rapid
  cracks}.
\newblock \emph{\bibinfo{journal}{Physical Review Letters}}
  \textbf{\bibinfo{volume}{114}}, \bibinfo{pages}{054301}
  (\bibinfo{year}{2015}).

\bibitem{sharon2001propagating}
\bibinfo{author}{Sharon, E.}, \bibinfo{author}{Cohen, G.} \&
  \bibinfo{author}{Fineberg, J.}
\newblock \bibinfo{title}{Propagating solitary waves along a rapidly moving
  crack front}.
\newblock \emph{\bibinfo{journal}{Nature}} \textbf{\bibinfo{volume}{410}},
  \bibinfo{pages}{68--71} (\bibinfo{year}{2001}).

\bibitem{ramanathan.97}
\bibinfo{author}{Ramanathan, S.} \& \bibinfo{author}{Fisher, D.~S.}
\newblock \bibinfo{title}{Dynamics and instabilities of planar tensile cracks
  in heterogeneous media}.
\newblock \emph{\bibinfo{journal}{Physical Review Letters}}
  \textbf{\bibinfo{volume}{79}}, \bibinfo{pages}{877--880}
  (\bibinfo{year}{1997}).

\bibitem{morrissey1998}
\bibinfo{author}{Morrissey, J.~W.} \& \bibinfo{author}{Rice, J.~R.}
\newblock \bibinfo{title}{Crack front waves}.
\newblock \emph{\bibinfo{journal}{Journal of the Mechanics and Physics of
  Solids}} \textbf{\bibinfo{volume}{46}}, \bibinfo{pages}{467--487}
  (\bibinfo{year}{1998}).

\end{thebibliography}

\end{document}